\documentclass{aa}  

\usepackage{graphicx}
\usepackage{txfonts}
\usepackage{graphicx}
\graphicspath{ {./plots/} }
\usepackage{hyperref}
\usepackage{placeins}
\usepackage{graphicx}

\usepackage{xcolor}
\usepackage{siunitx}

\usepackage{xspace}
\newcommand{\St}{\ensuremath{\mathrm{St}}\xspace}
\newcommand{\rc}{\ensuremath{r_\mathrm{c}}\xspace}

\begin{document}

   \title{Population Synthesis Models Indicate a Need for Early and Ubiquitous Disk Substructures}


   \author{Luca Delussu
          \inst{1},
          Tilman Birnstiel\inst{1,2},
           Anna Miotello \inst{3},
          Paola Pinilla \inst{4}, 
          Giovanni Rosotti\inst{5}, \and 
          Sean M. Andrews\inst{6}
          }

   \institute{University Observatory, Faculty of Physics, Ludwig-Maximilians-Universität München, Scheinerstr. 1, 81679 Munich, Germany
         \and
             Exzellenzcluster ORIGINS, Boltzmannstr. 2, D-85748 Garching, Germany
         \and European Southern Observatory, Karl-Schwarzschild-Strasse 2, 85748 Garching bei München, Germany
         \and Mullard Space Science Laboratory, University College London, Holmbury St Mary, Dorking, Surrey RH5 6NT, UK
         \and Dipartimento di Fisica ’Aldo Pontremoli’, Università degli Studi di Milano, via G. Celoria 16, I-20133 Milano, Italy
         \and Center for Astrophysics | Harvard \& Smithsonian, 60 Garden St., Cambridge, MA 02138, USA
             }

   \date{Received; accepted}

 
  \abstract
   {Large mm surveys of star forming regions enable the study of entire populations of planet-forming disks and reveal correlations between their observable properties. The ever-increasing number of these surveys has led to a flourishing of population study, a valuable tool and approach that is spreading in ever more fields. Population studies of disks have shown that the correlation between disk size and millimeter flux could be explained either through disks with strong substructure, or alternatively by the effects of radial inward drift of growing dust particles.}
   {This study aims to constrain the parameters and initial conditions of planet-forming disks and address the question of the need for the presence of substructures in disks and, if needed, their predicted characteristics, based on the large samples of disk sizes, millimeter fluxes, and spectral indices available.}
   {We performed a population synthesis of the continuum emission of disks, exploiting a two-population model (two-pop-py), considering the influence of viscous evolution, dust growth, fragmentation, and transport varying the initial conditions of the disk and substructure to find the best match to the observed distributions. Both disks without substructure and with substructure have been examined. We obtained the simulated population distribution for the disk sizes, millimeter fluxes and spectral indices by post-processing the resulting disk profiles (surface density, maximum grain size and disk temperature).}
   {We show that the observed distributions of spectral indices, sizes, and luminosities together can be best reproduced by disks with significant substructure, namely a perturbation strong enough to be able to trap particles, and that is formed early in the evolution of the disk, that is within \SI{0.4}{Myr}. Agreement is reached by relatively high initial disk masses ($10^{-2.3}M_{\star}\leqslant M_{disk}\leqslant10^{-0.5}M_{\star}$) and moderate levels of turbulence ($10^{-3.5}\leqslant\alpha\leqslant 10^{-2.5}$). Other disk parameters play a weaker role. Only opacities with high absorption efficiency can reproduce the observed spectral indices.}
   {Disk population synthesis is a precious tool for investigating and constraining parameters and initial conditions of planet-forming disks.
   The generally low observed spectral indices call for significant substructure, as planets in the Saturn to few Jupiter-mass range would induce, to be present already before \SI{0.4}{Myr}. Our results extend to the whole population that substructure is likely ubiquitous, so far assessed only in individual disks and implies that most "smooth" disks hide unresolved substructure.}

   \titlerunning{short title}
   \maketitle
%

\section{Introduction}
The advent of the Atacama Large Millimetre/Sub-Millimeter Array (ALMA) has strongly revolutionized the protoplanetary disk field. High-resolution observations have shown that most disks that have been imaged are characterized by the presence of substructures \citep{Andrews2018,Huang2018,Long2018}. Different mechanisms have been proposed to explain the origin of substructures in protoplanetary disks, such as planets \citep[e.g.,][]{Rice2006,Paardekooper2004,Pinilla2012a}, MHD processes \citep{Johansen2009,Bai2014}, binary companions \citep[e.g.]{Shi2012,Ragusa2017}, or variations in dust material properties \citep[e.g.,][]{Birnstiel2010,Okuzumi2016,Pinilla2017}. Nevertheless, it is widely believed that the presence of planets or protoplanets could be the reason behind the substructures observed in Class II disks, mainly due to kinematic evidence \citep{Teague2018,Pinte2018,izquierdo2022new} or directly imaged planets within the gaps, as in the case of PDS70 \citep{Muller2018,Keppler2018}. However, there is still a need to connect the distribution of exoplanets observed to a potential planet population in protoplanetary disks.

High-resolution observations enable studying and characterizing individual disks in great detail. However, these observations tend to be biased towards the brightest and largest disks. Nevertheless, ALMA has also been revolutionary by enabling large samples of lower-resolution observations. These provided hundreds of disks in large sample surveys of entire star-forming regions \citep[see][for a recent review]{Manara2023}. These observations have uncovered the existence of correlations between several disk-star observables such as the disk-size-luminosity relation \citep{Tripathi2017,Andrews2018a}, $M_\mathrm{dust}-M_\mathrm{star}$\citep{andrews2013mass,ansdell2016alma,pascucci2016steeper}, $\dot M - M_{dust}$ \citep{manara2016evidence} and the distribution of several key parameters of disks, such as the disk spectral index \citep{ricci2010dust_b,ricci2010dust_a,Tazzari2021}.

Surveys have thus opened up a new range of questions that cannot be answered by studying individual disks: what are the key mechanisms at play during the evolution of the disks that are needed to reproduce the distributions at different disk ages and the observed correlations? What mechanisms are responsible for the transport of angular momentum in disks (viscosity or MHD winds)? What are the initial conditions of the protoplanetary disks that are needed to reproduce observations? Surveys have thus led to the need to find models that can reproduce what we observe.
Disk population synthesis is the tool of choice to study these questions. Simulating thousands of models of protoplanetary disks evolving for several million years, disk population synthesis enables constraining disk initial conditions and identifying key evolutionary mechanisms of protoplanetary disks by comparing the results obtained from the simulations to the large amounts of surveys available.\\
\\
This study represents an extension of the study performed by \cite{zormpas2022large}. They explored whether smooth disks and sub-structured disks can reproduce the size-luminosity relation (SLR), performing a disk population study. In particular, they showed that smooth disks in the drift regime correctly reproduce the observed SLR \citep[see also][]{Rosotti2019}. However, their study has also shown that substructured disks can populate the bright part of the SLR, but they follow a different relation. Thus they state that the observed sample could be composed of a mixture of smooth and substructured disks. The performed disk population study has also exhibited the need to have a high initial disk mass and low turbulence to reproduce the observed distribution. Finally, they have shown that grain composition and porosity play a key role in the evolution of disks in the size-luminosity diagram. In particular, the opacity model from \cite{ricci2010dust_a} but with compact grains as in \cite{Rosotti2019} has proved to better reproduce the SLR than the DSHARP opacities (\citealp{Birnstiel2018}, which is similar in composition to previously used opacities such as in \citealp{Pollack1994} or \citealp{Dalessio2006}). A similar indication concerning DSHARP opacities comes from the study of \cite{stadler2022impact}, which shows that the opacity model of \cite{ricci2010dust_a} leads to a better matching of the observed spectral index in contrast to DSHARP opacities.\\
In this study, we aim to extend the work of \cite{zormpas2022large}, by exploring whether smooth and sub-structured disks can reproduce the observed spectral index distribution, and explore the required initial conditions that lead to evolved disks matching current observations. Moreover, we aim to understand whether it is possible to match both the size-luminosity and spectral index distributions at the same time and what initial disk parameters are required for that.\\
\\
This paper is structured as follows: in Sect. \ref{methods}, we describe our computational model for the evolution of the disk and introduce the analysis method used to compare to disk observations. Section \ref{results} introduces the main results obtained and the comparison to the observed distributions. We first focus on reproducing the spectral index and then on the possibility of a simultaneous matching of both the spectral index and the size-luminosity distributions. Section \ref{conclusions} presents our conclusions.

\section{Methods} \label{methods}
The two-population model (two-pop-py) by \cite{birnstiel2012simple} and  \cite{birnstiel2015dust} has been exploited to perform 1D simulations to describe the gas and dust evolution in the disk. Two-pop-py is a tool that is well suited for disk population studies since it captures the dust surface density evolution, the viscous evolution of the gas and the particle size with good accuracy. Being based on a set of simple equations, it allows to perform a single simulation quickly (order of seconds), making it computationally feasible to run large numbers of simulations within a reasonable amount of time. In the following we will describe the main characteristics of the two-pop-py model and the assumptions on which it is based.

\subsection{Disk evolution}
The protoplanetary gas disk is evolved according to the viscous disk evolution equation \citep{Lust1952,Lynden-Bell1974} using the turbulent effective viscosity as parameterized in \citep{shakura1973black},
\begin{equation} \label{viscosity_eq}
    \nu = \alpha_{gas} \frac{c_{s}^{2}}{\Omega_{k}} ;
\end{equation}
the dust diffusion coefficient is:
\begin{equation}
    D\ \simeq\ \alpha_{dust}\frac{c_{s}^{2}}{\Omega_{k}},
\end{equation}
where $\alpha_{gas}$, $c_{s}$ and $\Omega_{k}$ denote the turbulence parameter, the sound speed and the Keplerian frequency respectively. The additional term $\frac{1}{1+\St^{2}}$, with $\St$ being the Stokes number, as derived in \cite{youdin2007particle} was dropped since the Stokes number is always $< 1$ in all the simulations that have been performed.
\\
\\
As in \cite{zormpas2022large}, two different families of disk models have been taken into account in our study: \textit{smooth disks} and \textit{sub-structured disks}. A smooth disk is characterized by the absence of any gaps during its entire evolution while a sub-structured disk is a disk in which a single substructure is created during its evolution. In contrast to \cite{zormpas2022large} where substructure is inserted since the beginning of the evolution of the disk, our sub-structured disk starts as a smooth disk and during its evolution a gap is created, and thus the disk becomes sub-structured. This new approach allows investigating the effect of the time at which substructure is inserted during the evolution of the system. Substructure has been modelled as a gap due to the presence of a planet that is inserted in the disk. To mimic the presence of a planetary gap we have subdivided the $\alpha$ parameter into two different values: $\alpha_{gas}$ and $\alpha_{dust}$. The presence of the planet has been modelled as a local variation of the gas viscosity, while other than in the proximity of the gap, the two parameters are taken to be identical, that is $\alpha_{gas}=\alpha_{dust}$. More information on how we model the planet's presence in the disk can be found in Sect. \ref{planetgap_section}.
Adopting the two-population model described in \cite{birnstiel2012simple} we evolve the dust surface density assuming that the small dust is tightly coupled to the gas while the large dust particles can decouple from it and drift inward. 
The initial gas surface density follows the \cite{lynden1974evolution} self-similar solution,
\begin{equation}
    \Sigma_{g}(r) = \Sigma_{0} \left( \frac{r}{\rc} \right)^{-\gamma} exp \left[ - \left( \frac{r}{\rc} \right)^{2- \gamma} \right],
\end{equation}
where the normalization parameter $\Sigma_{0}=(2 - \gamma) M_{disk}/2\pi \rc^{2}$ is set by the initial disk mass $M_{disk}$, $\rc$ denotes the so-called characteristic radius of the disk and $\gamma$ the viscosity exponent. For our simulations, $\gamma$ has been set to $1$ for the initial profiles of all the disks. This choice is mostly consistent with our choice of the viscosity (see Eq. \ref{viscosity_eq}) which deviates from $\gamma = 1$ only in the case of substructure being included and in the isothermal part of the disk. Furthermore, choosing a $\gamma \neq 1$ would relax back to the $\gamma = 1$ case in a viscous time scale.
The initial dust surface density is related to the gas surface density by a constant initial dust-to-gas ratio $\Sigma_{d}/\Sigma_{g} = 0.01$.\\
\\
The two-population model consists of a population of initial grain size $a_{min}=0.1\mu m$ whose size is kept constant in time and space during the evolution and a large grain population that is allowed to increase in size with time. When \cite{ricci2010dust_a} opacities have been taken into account the particle bulk density has been set to a value of $\rho_{s} = \SI{1.7}{g/cm^3}$ as we considered the composition of \cite{ricci2010dust_a} but for the case of compact grains (no porosity), as in the model of \cite{Rosotti2019}. This opacity case will henceforth be referred to as \textit{Ricci compact opacities}. For DSHARP \citep{birnstiel2018disk} opacities we set $\rho_{s} = \SI{1.675}{g/cm^3}$ ($0\%$ porosity case) and this value decreases based on the porosity assumed for the grains. The grain composition consists of $60\%$ water ice, $30\%$ carbonaceous materials and $10\%$ silicates by volume. For DIANA \citep{woitke2016consistent} opacities we set $\rho_{s} = \SI{2.08}{g/cm^3}$. For a comparison with the size-luminosity distribution observed by \cite{Andrews2018a}, we computed the opacity in ALMA Band 7, more precisely at $0.89 mm$, and as described in Subsect. \ref{observables} we evaluated the continuum intensity profile of the disk considering the absorption and scattering opacity contributions (see Subsect. \ref{observables}). Four different grain porosity cases have been explored for the DSHARP opacity model \citep{birnstiel2018disk}: $0\%$, $10\%$, $50\%$ and $90\%$ porosity. The bulk densities adopted for these four cases are the following: $\rho_{s} = \SI{1.675}{g/cm^3}$, $\rho_{s} = \SI{1.508}{g/cm^3}$, $\rho_{s} = \SI{0.838}{g/cm^3}$ and $\rho_{s} = \SI{0.168}{g/cm^3}$. Ricci compact opacities have been adopted as the standard case in our analysis. When another model has been adopted it will be specified. For each simulation, the disk has evolved for 3 Myr.\\
\\
The 1D disk has been spatially modelled with a radial grid that ranges from 0.05 au to 2000 au, and the cells of the grid are spaced logarithmically. The main characteristics of the grid model are reported in Table \ref{table:1}.\\
\\
The temperature of the disk is linked to the luminosity of the star, that is we have worked within an adaptive temperature scenario. However, both the stellar luminosity $L_{\star}$ and effective temperature T are not evolved in our simulations. In appendix \ref{appendix:l_star} we show a comparison to the scenario in which $L_{\star}$, and thus T, evolve, showing that the results are analogous to the fixed scenario.\\
We have adopted a passive irradiated disk temperature model; no viscous heating or other processes have been considered. In particular, we have followed the temperature profile adopted by \cite{kenyon1996magnetic}:
\begin{equation}
    T = \left( \phi \frac{L_{\star}}{4 \pi \sigma_{SB}r^{2}} + (10K)^{4}\right)^{1/4} ,
\end{equation}
where $L_{\star}$ is the star luminosity, $\sigma_{SB}$ the Stefan-Boltzmann
constant and $\phi$ is the flaring angle set to $0.05$. The floor temperature of the disk has been set to $10K$.
The star's luminosity value $L_{\star}$ was set starting from the star's mass value $M_{\star}$ exploiting \cite{siess2000internet} evolutionary tracks and considering the value of $L_{\star}$ at an age of the star of 1Myr. $M_{\star}$ has been set based on the IMF of Chabrier and Kroupa (see next Sect. \ref{diskpop_section}).

\subsection{Disk Population synthesis}\label{diskpop_section}
Disk population synthesis is based on the idea of performing a large number of simulations of the evolution of dust and gas for millions of years, to constrain disk initial conditions and identify what are the key mechanisms at play in the disk by comparing the distribution of the observable parameters obtained from the simulations with the observed ones.
\begin{figure}
    \centering
    \includegraphics[scale=0.3]{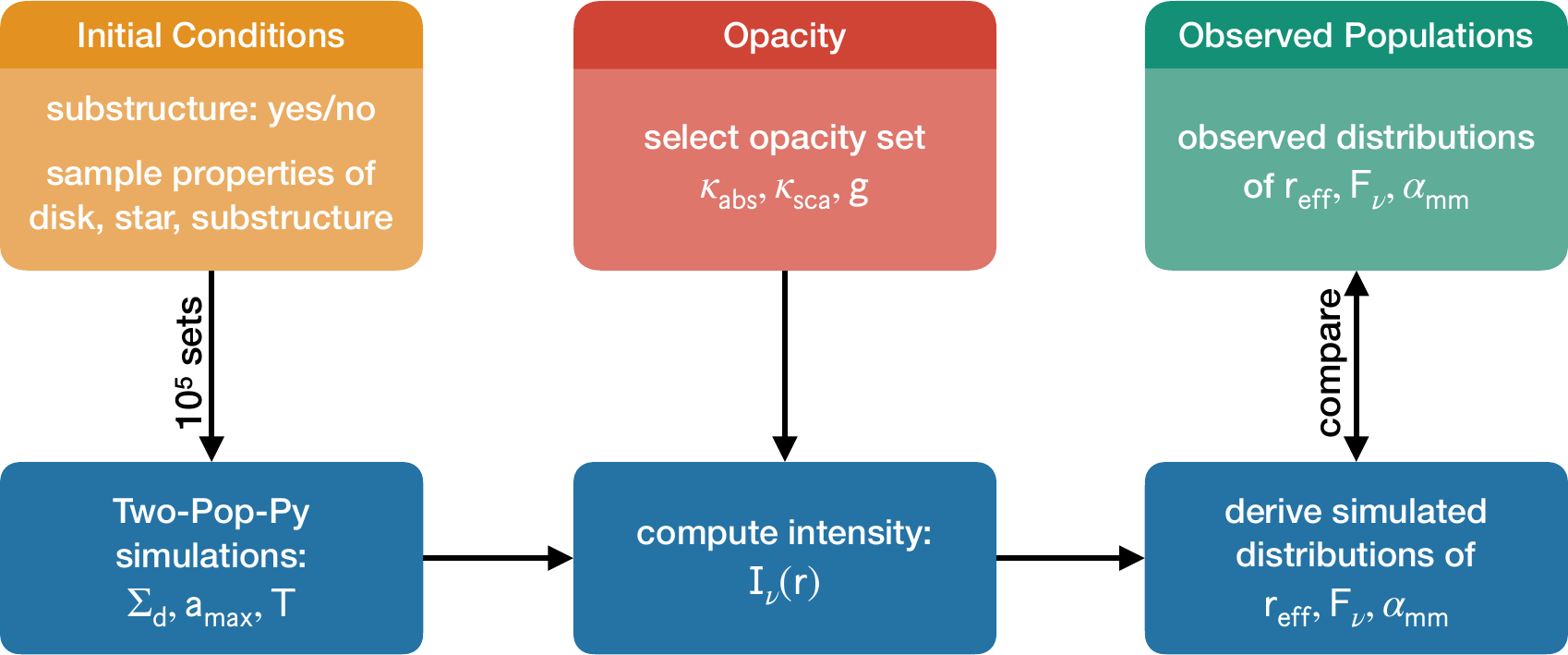}
    \caption{Disk population study procedure.}
    \label{schematic}
\end{figure}
\\
\\
Figure \ref{schematic} explains the key idea behind our disk population study. Firstly a large set of initial parameters has been set up combining different initial values of the main parameters used to describe each disk. In particular, to map all the parameter space, the set of initial conditions adopted for each disk has been constructed randomly drawing each parameter from a probability distribution function. A total of $10^{5}$ simulations have been performed for each population synthesis, which well map the entire relevant parameter space. The main parameters that have been taken into account to describe the disk are: disk mass ($M_{disk}$), stellar mass ($M_{\star}$), disk characteristic radius ($\rc$), viscous parameter ($\alpha$) and fragmentation velocity ($v_{frag}$). This choice of parameters applies to the description of all the disks (i.e., both smooth and sub-structured disks). Sub-structured disks have been characterized by three additional parameters: the mass of the planet that creates the gap ($m_{p}$), the time ($t_{p}$) and the position ($r_{p}$) at which the planet has been inserted. This set of three additional parameters applies for each planet that is inserted in the system, that is in the extra cases of double sub-structured systems (see Sect. \ref{extracases}), each gap has been characterized by its own set of three parameters ($m_{p,i}$, $t_{p,i}$, $r_{p,i}$). The drawing of the $m_{p}$ value is performed after drawing a value of $M_{disk}$, to impose the further physically reasonable restriction of $m_{p}<M_{disk}$. In the case of multiple substructures, we applied the following constraint: $\sum_{i} m_{p,i}<M_{disk}$.
Table \ref{table:2} shows the range that has been adopted for each parameter. 
\begin{table}
\tiny
\caption{Fixed parameters.}             
\label{table:1}      
\centering                          
\begin{tabular}{c c c}        
\hline\hline                 
Parameter & Description & Value or Range \\    
\hline                        
   $\Sigma_{d}/\Sigma_{g}$ & initial dust-to-gas ratio & 0.01 \\      
   $\rho_{s}\ [g/cm^{3}]$ & particle bulk density& 1.7 (Ricci opacity),\\
   &(no porosity)& 1.675 (DSHARP opacity)\\
   $\gamma$ & viscosity parameter & 1\\
   $r\ [au]$ & grid extent & 0.05-2000   \\
   $n_{r}\ [cells]$ & grid resolution & 400 \\ 
   $t\ [Myr]$ & duration of each simulation & 3 \\
\hline                                   
\end{tabular}
\end{table}

\begin{table}
\tiny
\caption{Disk initial parameters.}
\label{table:2}      
\centering                          
\scalebox{0.95}{
\begin{tabular}{c c c c}        
\hline\hline                 
Parameter & Description & Range & PDF \\    
\hline                        
 $\alpha$ & viscosity parameter & $10^{-4}-10^{-2}$ & log uniform\\
 $M_{disk}\ [M_{\star}]$ & initial disk mass & $10^{-3}-0.5$ & log uniform\\
 $M_{\star}\ [M_{\odot}]$ & stellar mass & $0.2-2.0$ & IMF\\
 $\rc\ [au]$ & characteristic radius & $10-230$ & log uniform\\
 $v_{frag}\ [cm/s]$ & fragmentation velocity & $200-2000$ & uniform\\
 $m_{p}\ [M_{\oplus}]$ & planet mass & $1-1050$ & uniform\\
 $r_{p}\ [\rc]$ & planet position & $0.05-1.5$ & uniform\\
 $t_{p}\ [Myr]$ & planet formation time & $0.1-0.4$ & uniform\\
 & & (early formation case) & \\
 $t_{p}\ [Myr]$ & planet formation time & $0.4-1.0$ & uniform\\
 & & (late formation case) & \\
\hline                                   
\end{tabular}}
\\
\tablefoot{Disk initial parameters and corresponding probability distribution function (PDF) from which their value is drawn for each single simulation. The drawing of the $m_{p}$ value was performed after the $M_{disk}$ to impose the further physically reasonable restriction of $m_{p}<M_{disk}$. In the case of multiple substructures, we applied the following constraint: $\sum_{i} m_{p,i}<M_{disk}$.
}
\end{table}
The viscosity parameter $\alpha$ and the initial mass of the disk $M_{disk}$ have been randomly drawn from a log uniform probability distribution function (PDF) to span uniformly the full range of values adopted for these two parameters.
A log uniform PDF has been adopted for the disk characteristic radius $\rc$ too, to favor smaller values of $\rc$ to resemble the observed behavior of protoplanetary disks' sizes. The values of the stellar mass $M_{\star}$ have been drawn from a functional form of the IMF proposed by \cite{maschberger2013function} which is based on the standard IMFs of Kroupa (\cite{kroupa2001variation},\cite{kroupa2002initial}) and Chabrier \citep{chabrier2003galactic}.\\
\\
Given the set of $10^{5}$ initial conditions, each disk has been evolved for $3 Myr$ using the two-pop-py evolutionary code. For the evolved disks obtained through two-pop-py, we have evaluated the observables parameters (Subsect. \ref{observables} and Subsect. \ref{spectral_methods}). Finally, the distributions of the observable parameters obtained for the simulated disks have been compared to the observed ones.\\
\\
Both observed and simulated fluxes of each disk have been scaled to a reference distance of 140pc.

\subsection{Planetary gaps}\label{planetgap_section}
The sub-structured disks have been produced generating a gap through the insertion of a planet during the evolution of the disk. The presence of a massive planet in the disk leads to the formation of a gap in the gas distribution, which we call substructure. To mimic the gap associated with a planet's presence in our disks, we have modified the value of the $\alpha_{gas}$ parameter. We took advantage of the inverse proportionality between $\alpha_{gas}$ and $\Sigma_{g}$ in a steady state regime, that is $\alpha_{gas} \propto 1/ \Sigma_{g}$. Thus, a bump in the $\alpha_{gas}$ profile will reflect in a gap in the $\Sigma_{g}$ profile that reproduces the presence of a planetary gap, besides, this procedure allows keeping a viscous evolution for $\Sigma_{g}$. We have adopted the \cite{kanagawa2016mass} prescription to model the gap created by a given planet at a given position in the disk. \\
The main parameter that describes the gap formed by the planet is:
\begin{equation} \label{kanagawa}
    K = \left( \frac{M_{p}}{M_{\star}} \right)^{2} \left( \frac{h_{p}}{R_{p}} \right)^{-5} \alpha^{-1} .
\end{equation}
The main caveat associated with the Kanagawa prescription is that it is an analytical approximation of the gap depth and width. However as described in \cite{zormpas2022large}, the width of the gap is the dominant factor in the evolution of the disk, and the depth is not crucial to the evolution of the disk. The position and radial extent of the gap are what matter the most to the evolution of the disk and those are well reproduced by the prescription.

\subsection{Observables}\label{observables}
One of the problems when dealing with protoplanetary disks is defining their size \citep[see][as a review]{miotello2023setting}. Indeed, as discussed in \cite{Tripathi2017} and \cite{rosotti2019time}, we cannot adopt the characteristic radius $\rc$ as a size indicator of the disks. We have thus followed the procedure of defining an effective radius $r_{eff}$ of the disk. The effective radius is defined as the radius that encompasses a given fraction of the total amount of flux that is produced by the disk. We have chosen to define our effective radius as the radius that encloses the $68\%$ of the total amount of flux produced by the disk, following \citet{Tripathi2017}.\\
The \cite{miyake1993effects} scattering solution of the radiative transfer equation has been adopted to evaluate the mean intensity $J_{\nu}$:
\begin{equation}
    \frac{J_{\nu}(\tau_{\nu})}{B_{\nu}(T(r))} = 1 - b \left( e^{-\sqrt{3\epsilon_{\nu}^{eff}}(\frac{1}{2}\Delta\tau -\tau_{\nu})} + e^{-\sqrt{3\epsilon_{\nu}^{eff}}(\frac{1}{2}\Delta\tau +\tau_{\nu})} \right),
\end{equation}
where $B_{\nu}$ is the Planck function and
\begin{equation}
    b = \left[ \left( 1 - \sqrt{\epsilon_{\nu}^{eff}} \right) e^{-\sqrt{3\epsilon_{\nu}^{eff}}\Delta\tau} + 1 + \sqrt{\epsilon_{\nu}^{eff}}\right]^{-1} ,
\end{equation}
and the optical depth $\tau_{\nu}$ is given by
\begin{equation}
    \tau_{\nu} = \left( k_{\nu}^{abs} + k_{\nu}^{sca,eff} \right) \Sigma_{d} ,
\end{equation}
where
\begin{equation}
    k_{\nu}^{sca,eff} = \left( 1 - g_{\nu} \right) k_{\nu}^{sca}
\end{equation}
is the effective scattering opacity and $k_{\nu}^{abs}$ is the dust absorption opacity, obtained from the Ricci compact \citep{Rosotti2019}, DSHARP \citep{birnstiel2018disk} or DIANA \citep{woitke2016consistent} opacities. $g_{\nu}$ is the forward scattering parameter. The introduction of the effective scattering opacity $k_{\nu}^{sca,eff}$ reduces the impact of the underlying approximation that scattering is isotropic. The effective absorption probability $\epsilon_{\nu}^{eff}$ is given by:
\begin{equation}
    \epsilon_{\nu}^{eff} = \frac{k_{\nu}^{abs}}{{k_{\nu}^{abs}} + k_{\nu}^{sca,eff}} ,
\end{equation}
and $\Delta\tau$ is
\begin{equation}
    \Delta\tau = \Sigma_{d}k_{\nu}^{tot}\Delta z .
\end{equation}
The intensity $I_{\nu}^{out}$ has been evaluated following the modified Eddington-Barbier approximation as adopted by \cite{zormpas2022large} following \cite{birnstiel2018disk}:
\begin{equation}
    I_{\nu}^{out} \simeq \left( 1 - e^{-\Delta\tau/\mu} \right) S_{\nu} \left( \frac{\Delta\tau}{2\mu} - \frac{2}{3} \right) ,
\end{equation}
with $\mu = cos\theta$ and the source function $S_{\nu}(\tau_{\nu})$ given by:
\begin{equation}
    S_{\nu}(\tau_{\nu}) = \epsilon_{\nu}^{eff}B_{\nu}(T_{d}) + (1 - \epsilon_{\nu}^{eff})J_{\nu}(\tau_{\nu}) .
\end{equation}
The role of scattering vanishes for small optical depth ($\Delta\tau<<1$). Indeed, the assumption that only the absorption opacity matters, is appropriate for optically thin dust layers of protoplanetary disks. However, one of the reasons behind the choice of including the scattering opacity in our treatment is that DSHARP survey \citep{birnstiel2018disk} revealed that optical depth is not small in protoplanetary disks. Moreover, \citep{kataoka2015millimeter} pointed out the importance of scattering in the (sub-)millimeter polarization of protoplanetary disks.\\ \\
A particle size distribution, evolved by two-pop-py, is needed to compute the optical properties of the dust. We assumed a population of grains described by a power-law size distribution, $n(a) \propto a^{-q}$, with q=2.5 for $a_{min}\leqslant a \leqslant a_{max}$. The grain size at each radius is set by the lower value among the maximum grain size possible in the fragmentation- or drift-limited regimes, As described in \cite{birnstiel2012simple}, disks in the drift-limited regime are better described by $q=2.5$, while disks in the fragmentation-limited regime by $q=3.5$. Following the same reasoning reported in \cite{zormpas2022large}, we adopt a value of $q=2.5$. Indeed smooth disks are mostly drift-limited and if there is a fragmentation-limited region in the disk, it resides in the inner part of the disk, thus the luminosity of the disk will still mainly depend on the drift region as it resides in the external part of the disk where there is the bulk of the disk mass. The fragmentation-limited region can be located further out in sub-structured disks in correspondence to the ring, but the difference between the choice of the two $q$ exponents is reduced by the fact that the rings are mostly optically thick.

\subsection{Spectral index}\label{spectral_methods}
We define the spectral index as the slope of the (sub-)mm SED of the dust emission, that is:
\begin{equation}
    \alpha_{mm} = \frac{dlogF_{\nu}}{dlog \nu},
\end{equation}
where $F_{\lambda}$ is the \textit{disk-integrated flux} at a given wavelength $\lambda$, thus $\alpha_{mm}$ is the \textit{disk-integrated spectral index}.  
Since we usually deal with frequencies that are very close to each other we can write:
\begin{equation}
    \alpha_{mm} = \frac{log(F_{\lambda,1}/F_{\lambda,2})}{log(\lambda_{1}/\lambda_{2})} .
    \label{spectral_eq}
\end{equation}
If we assume that the radiation is emitted in a Rayleigh-Jeans regime and that the emission is optically thin we can further relate, in first approximation, the spectral index to the dust opacity power law slope $\beta$ ($k_{\nu} \propto \nu^{\beta}$):
\begin{equation}
    \alpha_{mm} \approx \beta +2 .
\end{equation}
The spectral index represents a key parameter for the characterization of protoplanetary disks because it carries information about the maximum size of the particles that are present in the disk (\cite{miyake1993effects},\cite{natta2004star},\cite{draine2006submillimeter}).\\
\\
Starting from the post-processed values of the fluxes that have been obtained for each disk at different wavelengths, we have evaluated the spectral index of each simulated disk at different snapshots of their evolution. In particular, since we have referenced to the work of \cite{tazzari2021first}, we have considered $\lambda_{2}=\SI{0.89}{mm}$ and $\lambda_{1}=\SI{3.10}{mm}$ and we have applied Eq. \ref{spectral_eq} to determine the spectral index.\\
\section{Results} \label{results}
The following section contains the main results obtained through our analysis. Subsection \ref{spectral_index} firstly introduces the results obtained for the reproducibility of the spectral index distribution and then focuses on the simultaneous reproducibility of both the spectral index and the size-luminosity distribution. We underline the characteristics of the disks needed to match the observed distributions. In Subsect. \ref{extracases} we extend the analysis to some extra cases (different opacity models, IMF and number of substructures). In Subsect. \ref{future_subsection} we present some extra results that are linked to future development of this work and open problems in the protoplanetary disk field. To compare the simulated distributions with the observed ones, a possible age spread of the simulated disks was taken into account. That is, for each simulated disk, the observables taken into account for the creation of the overall simulated distributions were randomly selected from the snapshots at \SI{1}{Myr}, \SI{2}{Myr}, and \SI{3}{Myr}.\\
We compare our simulated spectral index distributions to the observed sample adopted in \cite{tazzari2021first}. The latter is a collection of disks from Lupus region, detected at 0.89mm \citep{ansdell2016alma} and 3.1mm \citep{tazzari2021first}, and Taurus and Ophiucus star-forming regions \citep{ricci2010dust_b,ricci2010dust_a}. We compare our simulated size-luminosity distribution to the observed sample reported in \cite{Andrews2018a}.

\begin{figure*}[!htb]
    \centering
    \includegraphics[scale=0.45]{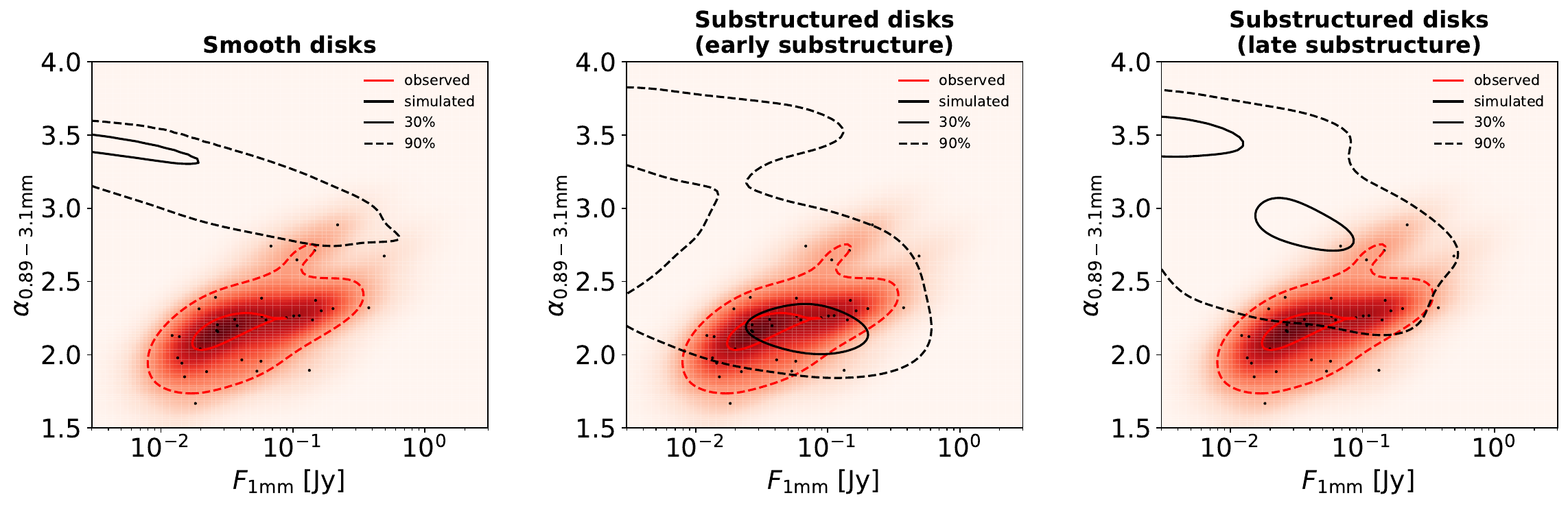}
    \caption{Spectral index distribution smooth disks (left) vs sub-structured disks (middle and right), for the entire parameter space of initial conditions (Table \ref{table:2}). \\Middle plot: planet randomly inserted in a range between 0.1-0.4 Myr. Right plot: planet randomly inserted in a range between 0.4-1 Myr.\\ Heatmap of the observed disks with the black dots representing each single observed disk. The black and red lines refer to the simulated results and the observational results respectively. In particular, the continuous lines encompass the $30\%$ of the cumulative sum of the disks produced from the simulations or observed. The dashed lines encompass the $90\%$ instead. \\ We compare our simulated spectral index distribution to the observed sample adopted in \cite{tazzari2021first}. The latter is a collection of disks from Lupus region, detected at 0.89mm \citep{ansdell2016alma} and 3.1mm \citep{tazzari2021first}, and Taurus and Ophiucus star-forming regions \citep{ricci2010dust_b,ricci2010dust_a}.}
    \label{smooth_vs_substr_earlylate}
\end{figure*}

\subsection{Spectral index and size-luminosity distribution} \label{spectral_index}
\begin{figure*}[!tbh]
    \centering
    \includegraphics[scale=0.6]{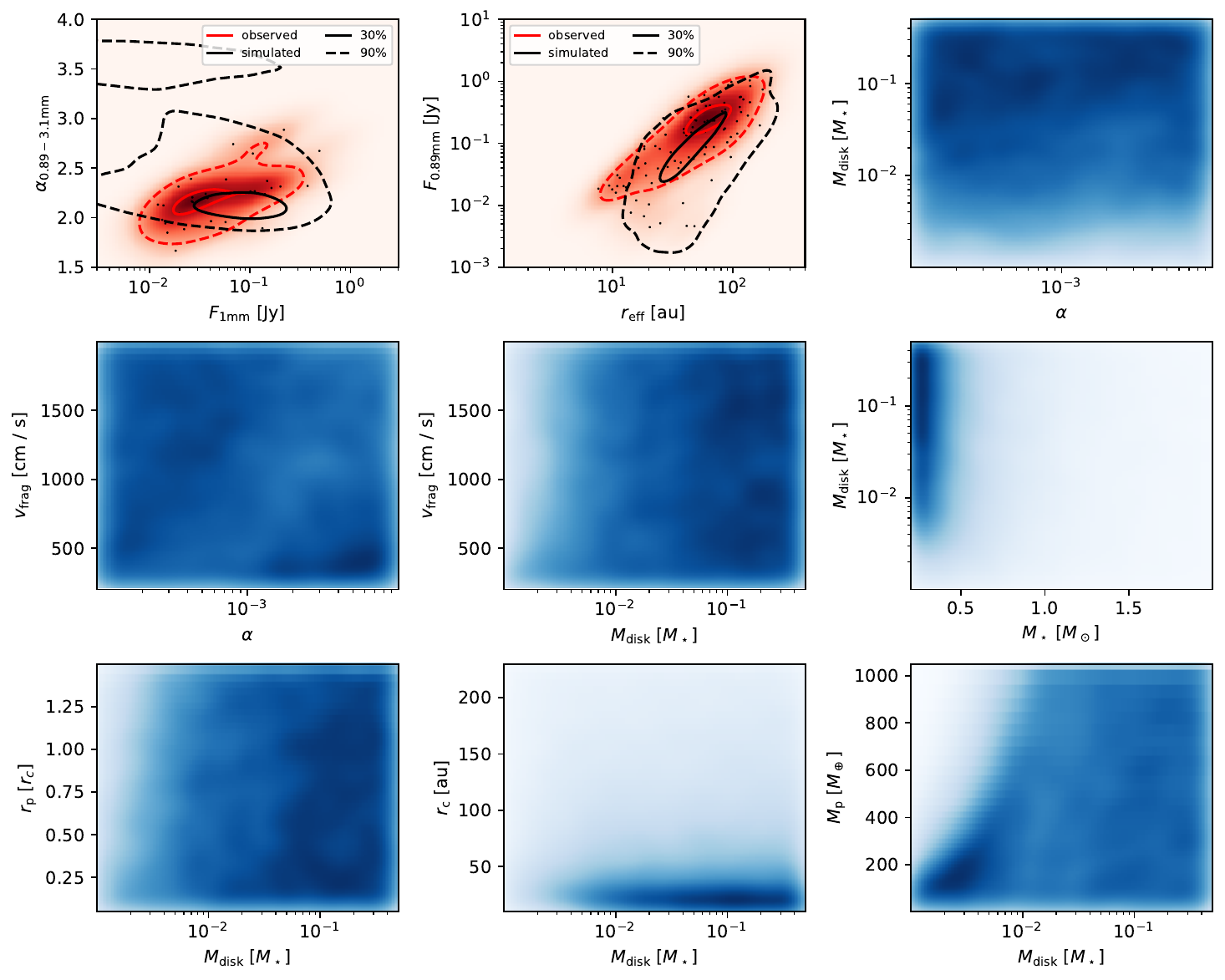}
    \caption{Spectral index, size-luminosity and initial parameters distributions for the parameter space of the initial conditions selecting disks with a spectral index $0\lnsim \alpha_{0.89-3.1mm}\lnsim4$, $10^{-3}Jy\lnsim F_{1mm}\lnsim10Jy$, $10^{-3}Jy\lnsim F_{0.89mm}\lnsim10Jy$ and $10^{0.1}au\lnsim r_{eff}\lnsim10^{2.6}au$. We compare our simulated spectral index distribution to the observed sample adopted in \cite{tazzari2021first}. The latter is a collection of disks from Lupus region, detected at 0.89mm \citep{ansdell2016alma} and 3.1mm \citep{tazzari2021first}, and Taurus and Ophiucus star-forming regions \citep{ricci2010dust_b,ricci2010dust_a}. We compare our simulated size-luminosity distribution to the observed sample reported in \cite{Andrews2018a}.}
    \label{allphasespace}
\end{figure*}

\begin{figure}[!htb]
    \centering
    \includegraphics[scale=0.35]{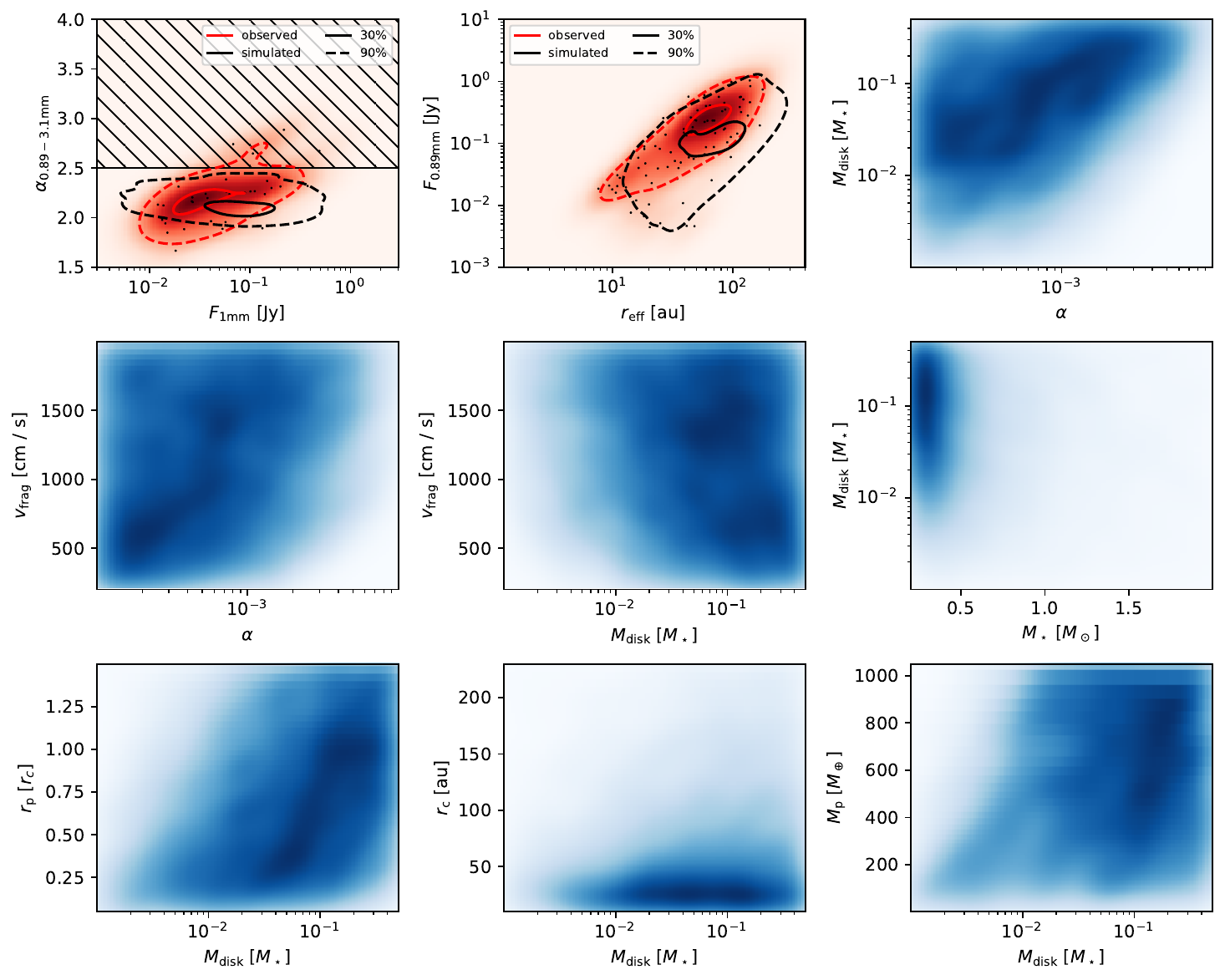}
    \caption{Spectral index, size-luminosity and initial parameters distributions selecting disks with a spectral index $\alpha_{0.89-3.1 mm}\leq2.5$.}
    \label{spectralbelow25}
\end{figure}

\begin{figure}[htb]
    \centering
    \includegraphics[scale=0.35]{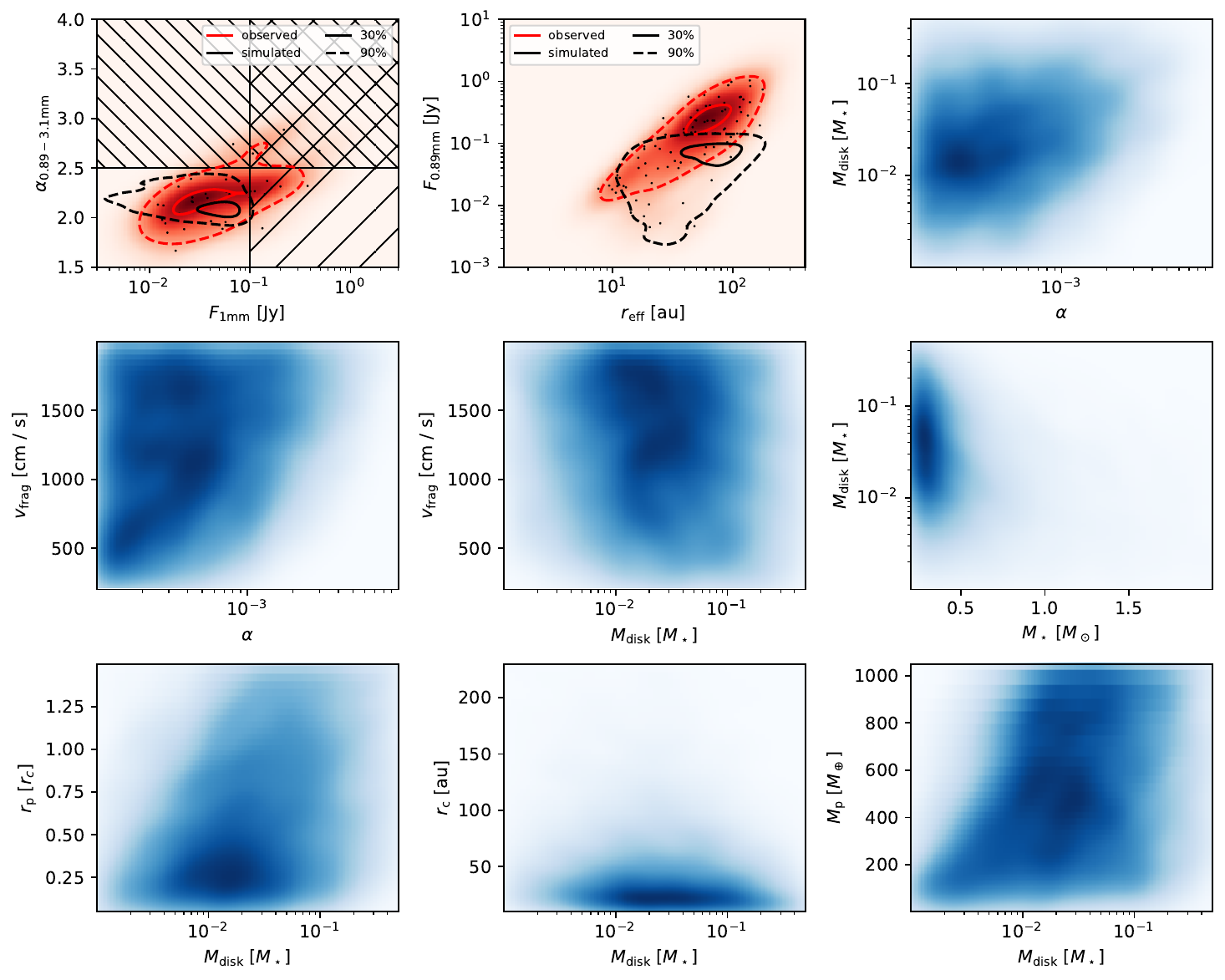}
    \caption{Spectral index, size-luminosity and initial parameters distributions selecting disks with a spectral index $\alpha_{0.89-3.1 mm}\leq2.5$ and a flux $F_{1mm}\leq0.1Jy$.}
    \label{spectral_leftregion}
\end{figure}

\begin{figure}[htb]
    \centering
    \includegraphics[scale=0.35]{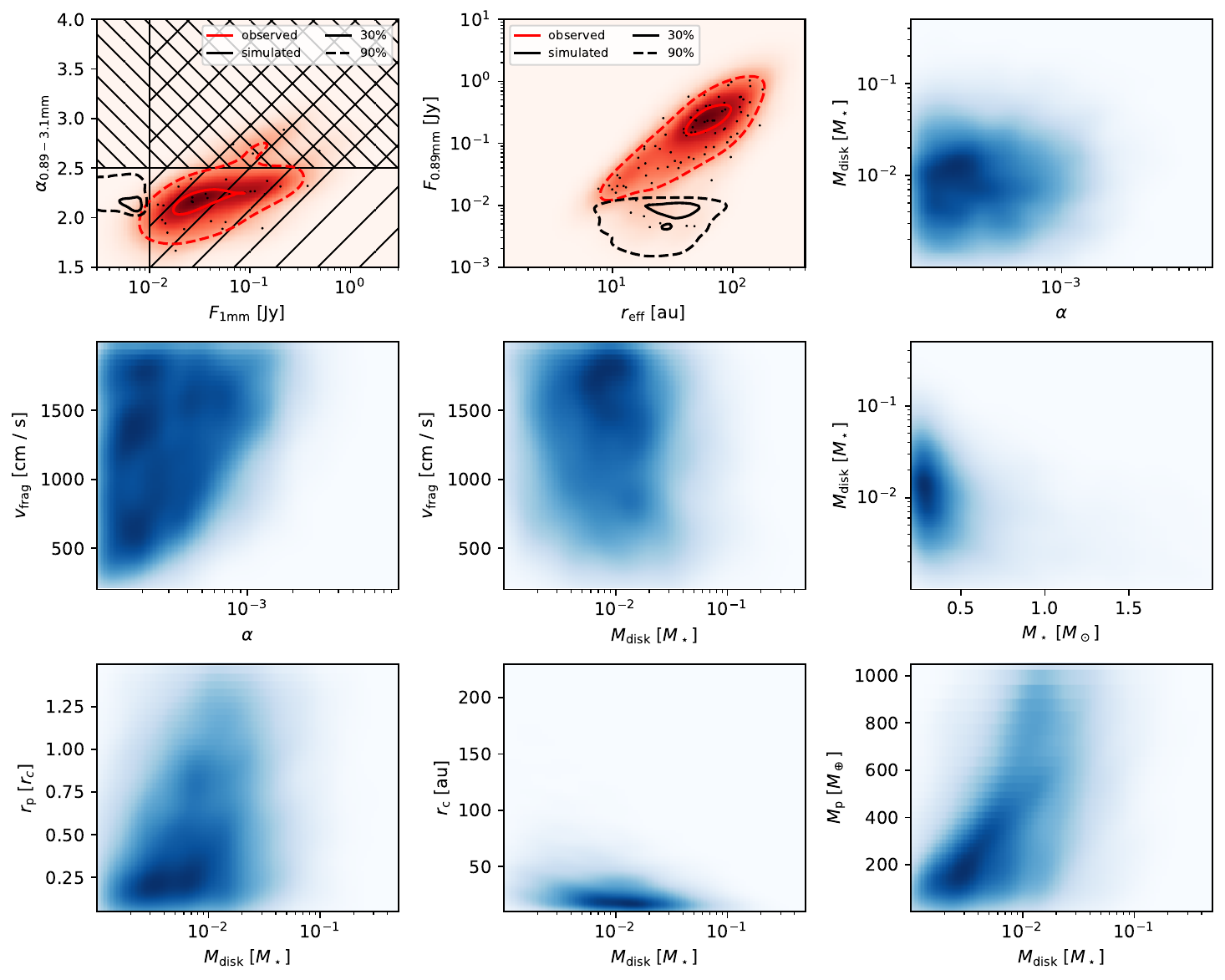}
    \caption{Spectral index, size-luminosity and initial parameters distributions selecting disks with a spectral index $\alpha_{0.89-3.1 mm}\leq2.5$ and a flux $F_{1mm}\leq0.01Jy$.}
    \label{spectral_lowalpha}
\end{figure}

\begin{figure}[htb]
    \centering
    \includegraphics[scale=0.35]{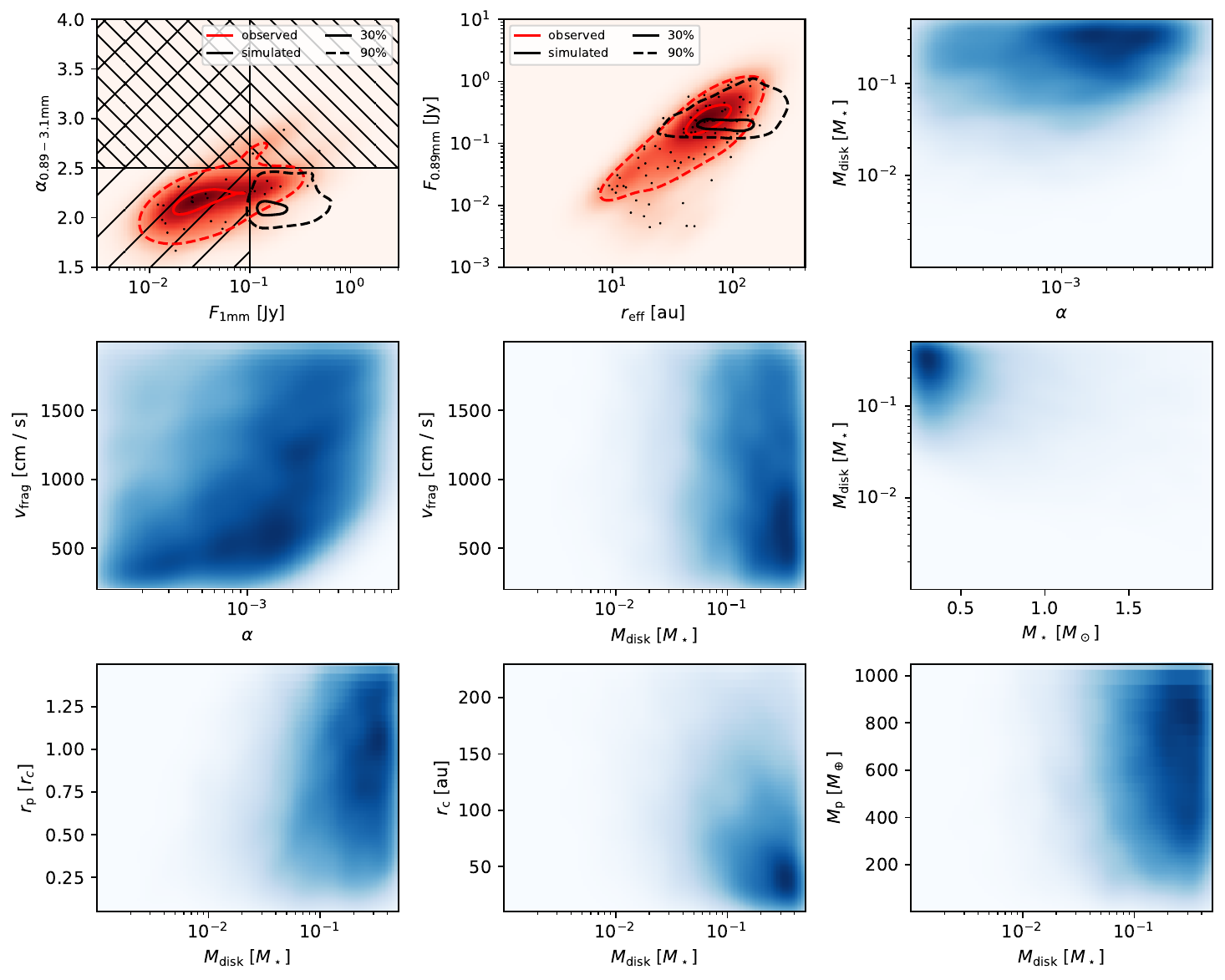}
    \caption{Spectral index, size-luminosity and initial parameters distributions selecting disks with a spectral index $\alpha_{0.89-3.1 mm}\leq2.5$ and a flux $F_{1mm}>0.1Jy$.}
    \label{spectral_rightregion}
\end{figure}

\begin{figure*}[!htb]
    \centering
    \includegraphics[scale=0.6]{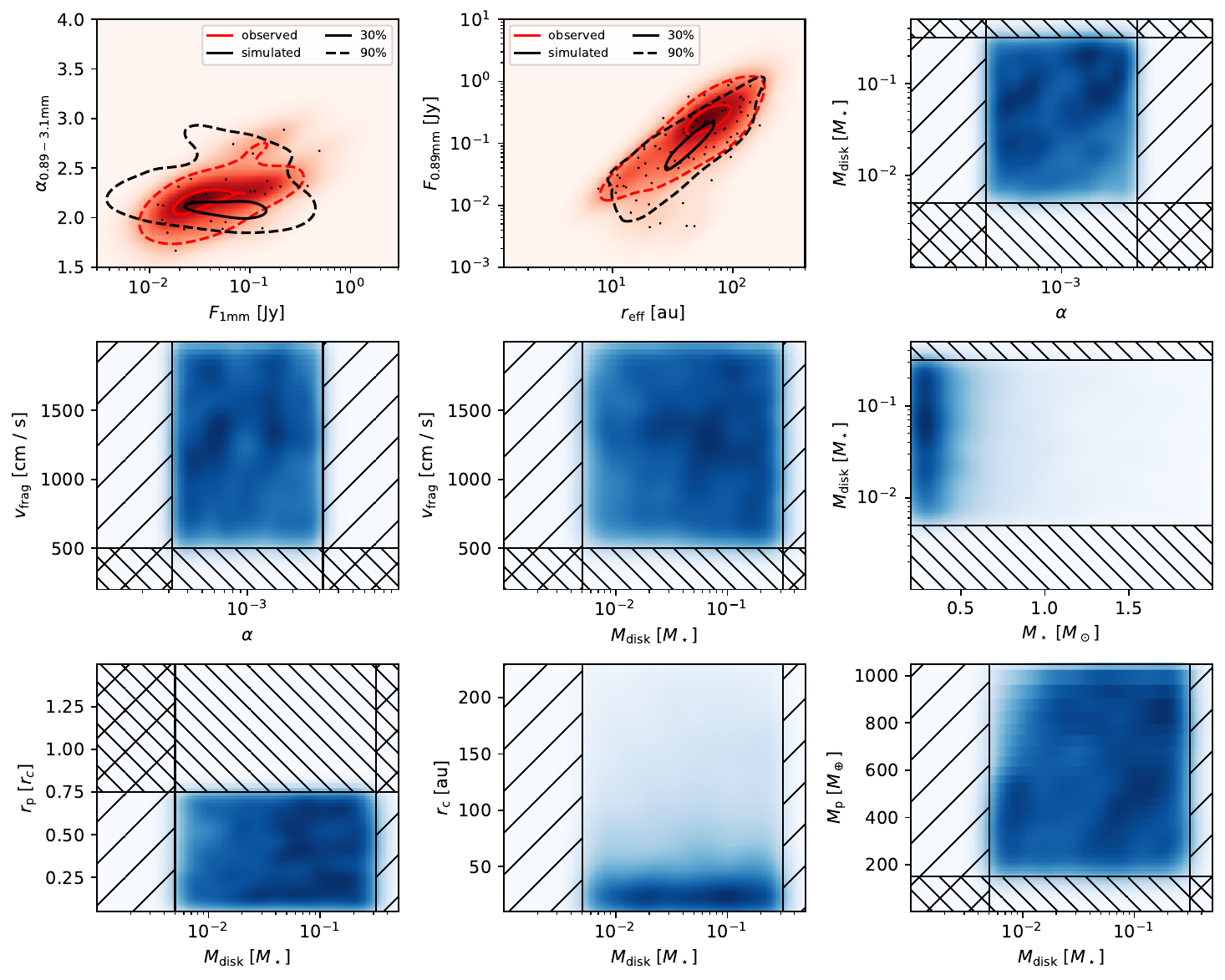}
    \caption{Spectral index, size-luminosity and initial parameters distributions selecting disks with: $10^{-3.5}\leqslant\alpha\leqslant 10^{-2.5}$, $10^{-2.3}M_{\star}\leqslant M_{disk}\leqslant10^{-0.5}M_{\star}$, $v_{frag}\geq500 cm/s$, $m_{p}\geq150M_{\oplus}$, $r_{p}\leq0.75\rc$.}
    \label{spectral_finalcut}
\end{figure*}

\begin{figure}[!htb]
    \centering
    \includegraphics[scale=0.45]{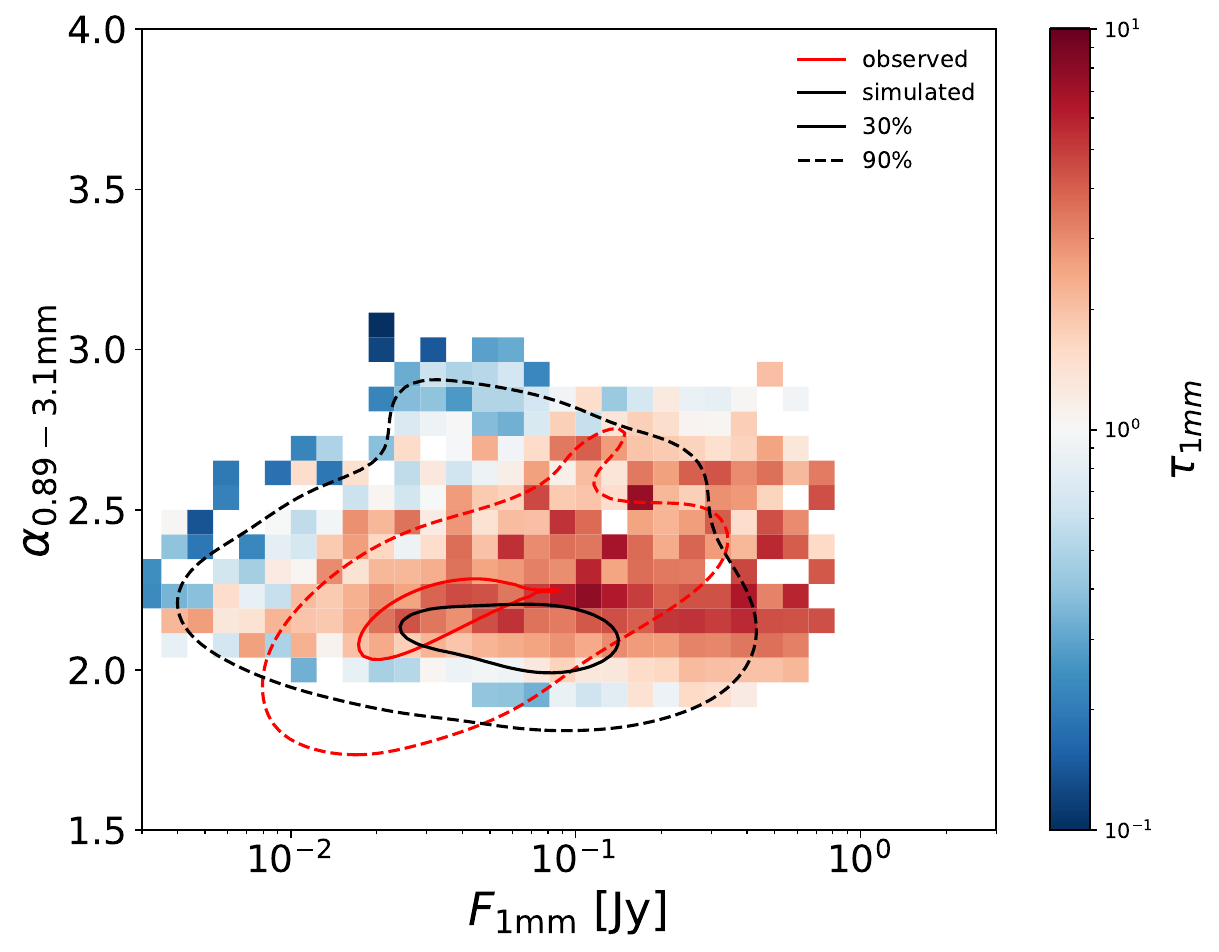}
    \caption{Spectral index and flux averaged optical depth distribution at 1mm ($\tau_{1mm}$), selecting disks with: $10^{-3.5}\leqslant\alpha\leqslant 10^{-2.5}$, $10^{-2.3}M_{\star}\leqslant M_{disk}\leqslant10^{-0.5}M_{\star}$, $v_{frag}\geq500 cm/s$, $m_{p}\geq150M_{\oplus}$, $r_{p}\leq0.75\rc$. The optical depth associated with each bin is the mean of the flux averaged optical depths of the disks belonging to each bin.}
    \label{opticaldepth}
\end{figure}

The main focus of this study has been to investigate the population synthesis outcome for the spectral index distribution of smooth and sub-structured disks. The left and middle panels of Fig. \ref{smooth_vs_substr_earlylate} show the clear difference that has been found between smooth and sub-structured disks. Indeed, if we consider the entire parameter space of the initial conditions adopted and reported in Table \ref{table:2}, we observe the first important indication obtained through our simulations: smooth disks cannot produce a low value of the spectral index, that is $\alpha_{0.89-3.1 mm}\leq2.5$, while sub-structured disks produce both spectral index values around 3.5 and below 2.5, thus populating the observed spectral index region. This result confirms and extends the findings of previous studies \cite[e.g][]{pinilla2012trapping} that had already shown the need for the presence of substructures to reproduce the spectral index values observed modeling individual disks.\\
\\
This is because a small value of the spectral index could be achieved due to the production of large dust particles and/or to the presence of optically thick regions originating from the accumulation of large quantities of solids. Neither can be achieved in a smooth disk due to the radial drift mechanism which preferentially removes the largest grains and depletes the overall dust mass. Moreover, since radial drift is stronger for the largest particles, it not only avoids creating more massive particles in general but also removes the massive particles created in the disk even faster. Substructure stops or slows down particle radial drift enabling particle growth and/or leading to the formation of optically thick regions. An in-depth investigation of the real cause behind the production of low spectral index values in the case of the presence of substructure is provided later in this section.\\
\\
As a next step, we derive the initial conditions of the substructured disks that allow the production of a population in agreement with the observed one. In this regard, a second important result uncovered by this analysis involves the formation time of substructure. The middle and right panels of Fig. \ref{smooth_vs_substr_earlylate} show the comparison between the spectral index distribution obtained for disks in which the substructure is inserted in a range between \SI{0.1}{Myr} and \SI{0.4}{Myr} after the start of the simulation (early formation case) and those in which it is inserted in a range between \SI{0.4}{Myr} and \SI{1}{Myr} (late formation case). The delayed insertion of the substructure leads to the production of spectral indices tending to higher values, larger than those observed. Thus, both the ubiquity of substructures and their rapid formation are required to produce spectral index values in the observed range.\\
\\
If the substructure is thought to be caused by a planet, as in our case, such a constraint on the formation time of substructure translates into an indication of the formation time of the planet associated with it. This leads to important insights into planetary formation theories. Namely, that planet formation is fast, in contrast with earlier formation models, such as \cite{pollack1996}. The result obtained on the rapid formation of substructure appears to be along the same lines as studies concerning the formation of giant planets such as \cite{savvidou2023make}, and with the results obtained by \cite{stadler2022impact} when exploring the possibility of producing a small spectral index for different sets of initial disk parameters.\\
\\
Having ascertained these two main results (ubiquity
of substructures and their rapid formation), we therefore focused on the study of disks with substructure and in which they form rapidly ($0.1 Myr \leqslant t_{p} \leqslant 0.4 Myr$).\\
\\
To characterize the initial conditions necessary to reproduce the observed distribution of the spectral index, the distribution of initial parameters associated with different selected regions in the spectral index vs. flux diagram was analyzed. Figure \ref{allphasespace} shows the comparison between the observed and simulated distribution of the spectral index (top left panel) and the size luminosity distribution (top middle panel), while the blue panels show the distribution of parameters sampled. In particular, red contours represent the distribution we want to match while the black contours in the observational space show the resulting distribution of our population. Figure \ref{initialcond_noobscuts} in Appendix \ref{appendix:initial_conditions} shows the distribution of the initial parameters for the entire sampled parameter space (see Table \ref{table:2}). However, to compare the simulated disks with the observed ones, we filtered our original dataset selecting disks with a spectral index, size and flux of the order of the observed ones (see Fig. \ref{allphasespace} and its description).\\
Figure \ref{allphasespace} shows that we need a $M_{disk}\gtrsim10^{-2.3}M_{\star}$ to produce disks with a flux of the order of the observed disks.\\
\\
Figures \ref{spectralbelow25}, \ref{spectral_leftregion}, \ref{spectral_lowalpha} and \ref{spectral_rightregion} show the initial conditions and their distribution for three different regions in the spectral index vs. flux diagram. Figure \ref{spectralbelow25} shows the distribution for the initial conditions leading to disks with a spectral index below 2.5. Figure \ref{spectral_leftregion}, Fig. \ref{spectral_lowalpha} and Fig. \ref{spectral_rightregion} show the initial conditions that lead to the production of disks populating three different flux regions ($F_{1mm}\leq0.1Jy$, $F_{1mm}\leq0.01Jy$ and $F_{1mm}>0.1Jy$, respectively), still in the case of disks producing a spectral index below 2.5.\\
\\
Figure \ref{spectralbelow25} shows a relationship between $\alpha$ and $M_{disk}$; to produce disks with a spectral index below 2.5, as $\alpha$ increases, $M_{disk}$ must also increase. This general trend between $\alpha$ and $M_{disk}$ primarily reflects the fact that to produce a low spectral index value, it is necessary to have an efficient trapping mechanism. For low $\alpha$ values, the trapping mechanism is highly efficient, and thus even disks that are not extremely massive are capable of producing low spectral index values, since although the system has little material available, it is still able to trap most of it. As $\alpha$ increases, the efficiency of the trapping mechanism tends to decrease, for example increased diffusivity, that is more grains will escape the bump, and decreased particle size in the fragmentation limit which are less efficiently trapped by the radial drift \citep{zhu2012dust}, so to cope with this effect it is necessary to increase the reservoir of material available. Most of the material will not be trapped due to the lower trapping efficiency, but there will be enough material available for a good amount to be trapped and produce a low spectral index value. The observed trend also partly reflects the condition imposed on the drawing of the $m_{p}$ value (i.e., $m_{p}<M_{disk}$). Indeed, Eq. \ref{kanagawa}, which describes the efficiency of the trapping mechanism, shows that low $\alpha$ values favor the trapping mechanism and generally make it unnecessary to have corresponding high $m_{p}$ values. Nevertheless, to keep the trapping mechanism efficient as $\alpha$ increases, the value of $m_{p}$ must be increased. But since the latter is limited by the condition $m_{p}<M_{disk}$, it is necessary to increase the value of $M_{disk}$ to have access to the desired production of more massive planets.\\
\\
In more detail, as shown in Fig. \ref{spectral_leftregion}, if we also add a condition on the final flux value, we see that this flux region is populated by disks having $10^{-2.3}M_{\star}\lnsim M_{disk}\lnsim10^{-1}M_{\star}$ and a value of $\alpha\lnsim10^{-3}$ in the $M_{disk}$ vs $\alpha$ space; the same upper limit on $\alpha$ is observed in the $v_{frag}$ vs $\alpha$ space. The upper limit for $M_{disk}$ is simply related to the fact that we want to populate a low-flux region. The upper limit for $\alpha$ comes from the fact that wanting to populate a low flux region we need lighter disks, thus a smaller value of $\alpha$ because we need to trap the little material available efficiently. Furthermore, a selection of disks with $v_{frag}\gtrsim500 cm/s$ in the $v_{frag}$ vs $M_{disk}$ space can be noted. Indeed for values of $v_{frag}\lesssim500 cm/s$ it becomes harder to trap as the Stokes number will become too small. It is further noted that this region is mostly populated by disks with $r_{p}\lnsim0.75\rc$. Indeed, placing the gap too far away will lead to the production of a higher flux as the surface area of the ring will increase. More in detail, Fig \ref{spectral_lowalpha} shows that disk with a flux $F_{1mm}\leq0.01Jy$, lower than the observed disks' fluxes, are associated with the lowest $\alpha$ values, i.e $\alpha\lesssim 10^{-3.5}$.\\
In contrast, Fig. \ref{spectral_rightregion} shows that the "high-flux" region ($F_{1mm}>0.1Jy$) is characterised by disks with an $\alpha$ that covers the full range $10^{-4}-10^{-2}$ in the $M_{disk}$ vs $\alpha$ space, but favouring the $\alpha\gtrsim10^{-3}$ cases, and in general a $M_{disk}\gtrsim10^{-1}M_{\star}$. The latter result reflects the request to populate the "high-flux" region. A relationship between $v_{frag}$ and $\alpha$ is also evident, as one grows, so does the other. There is also a selection for planets with $m_{p}\gtrsim150M_{\oplus}$ as in this case we are dealing with higher $\alpha$ in general, so a higher mass of the planet is needed to increase the efficiency of the trapping.
\\
\\
Having obtained these constraints on the initial conditions, we can now argue in the other direction: which cuts need to be imposed on the initial conditions to obtain a match between the simulated and observed distribution. Figure \ref{spectral_finalcut} shows the result obtained by selecting the disks associated to the following initial conditions, for the early formation scenario (i.e., $0.1 Myr \leqslant t_{p} \leqslant 0.4 Myr$): $10^{-3.5}\leqslant \alpha \leqslant 10^{-2.5}$, $10^{-2.3}M_{\star}\leqslant M_{disk}\leqslant10^{-0.5}M_{\star}$, $v_{frag}\geq500 cm/s$, $m_{p}\geq150M_{\oplus}$ and $r_{p}\leq0.75\rc$. By filtering these disks, it is possible to obtain a spectral index distribution consistent with that observed. While more sophisticated constraints might lead to an even more precise match, the purpose of this work is to show that it is possible to reproduce the observed distribution of the spectral index from a disk population synthesis and outline the most basic constraints on the initial conditions necessary to achieve this.\\
Although the cuts made were based solely on an analysis of the behavior of the spectral index distribution, Fig. \ref{spectral_finalcut} shows a remarkable result, that is not only applied cuts allow reproducing the spectral index distribution, but they simultaneously yield a distribution in the size-luminosity diagram that matches the observed one.\\
\\
Figure \ref{slr_bottom} and Fig. \ref{slr_top} reported in Appendix \ref{appendix:slr_analysis} show the distribution of initial parameters associated with two different regions of the size-luminosity diagram. Both regions exhibit a spread in the initial $\alpha$ value that, while favoring intermediate (low-flux region) or high-intermediate (high-flux region) values, nevertheless extends over the entire range. Similarly to the two regions previously analyzed for the spectral index, in the size-luminosity diagram, the low-flux region requires $M_{disk}\gtrsim10^{-2.3}M_{\star}$, while the high-flux region becomes more selective as it requires $M_{disk}\gtrsim10^{-1}M_{\star}$. Furthermore, in the case of the low-flux region, the selection of $r_{p}\leq0.75\rc$ is observed again. It is therefore reasonable that the cuts applied to reproduce the observed distribution of the spectral index also lead to an automatic and simultaneous matching of the distribution in the size-luminosity diagram. Besides, no further cuts appear necessary to improve the matchings. Better matchings could be obtained only by refining the way the cuts are made, that is imposing non-homogeneous distributions for the initial conditions and introducing an analytical way to compare the simulated disks' distribution to the observed one.\\
\\
An in-depth investigation of the real cause behind the production of low spectral index values for substructured disks has been performed evaluating the flux averaged optical depth associated to each simulated disk. Figure \ref{opticaldepth} shows the value of the flux averaged optical depth obtained for the distribution of substructured disks obtained applying our best cuts. The figure reveals that a small value of the spectral index is achieved by the production of an optically thick region in the disk, originating from the accumulation of material due to the presence of substructure. We can also notice the presence of optically thin or marginally optically thin disks, which explains why we can now reproduce the size-luminosity distribution with only substructured disks. Indeed, \cite{zormpas2022large} study, from which it emerged that a mix of smooth and substructured disks was necessary to reproduce the SLR, relied on the hypothesis that all substructured disks were optically thick. The different parameter space adopted in our study, in particular the choice to draw $M_{\star}$ from the IMF instead of a uniform distribution, leads to a population of disks composed of a mix of optically thick substructured disks and optically thin (or marginally optically thin) substructured disks. This reflects directly on the final distribution of the simulated disks in the size-luminosity diagram. While massive disks, populating the high flux region ($-1\leqslant log\ F_{0.89mm}[Jy]\leqslant0.5$) of the size-luminosity diagram, will still populate the observed region as they are optically thick as in \cite{zormpas2022large}; disks in the low flux region ($-2\leqslant log\ F_{0.89mm}[Jy]\leqslant-1$) are now optically thin or marginally optically thin, and not optically thick as assumed in \cite{zormpas2022large}, thus they experience a change (i.e., reduction) in their flux allowing them to fall in the observed region. Thus we can now reproduce the observed size-luminosity distribution having no longer required a mix of smooth and substructured disks.

\clearpage
\subsection{Case studies on opacities, the IMF and double substructure} \label{extracases}
\begin{figure*}[htb]
    \centering
    \includegraphics[scale=0.45]{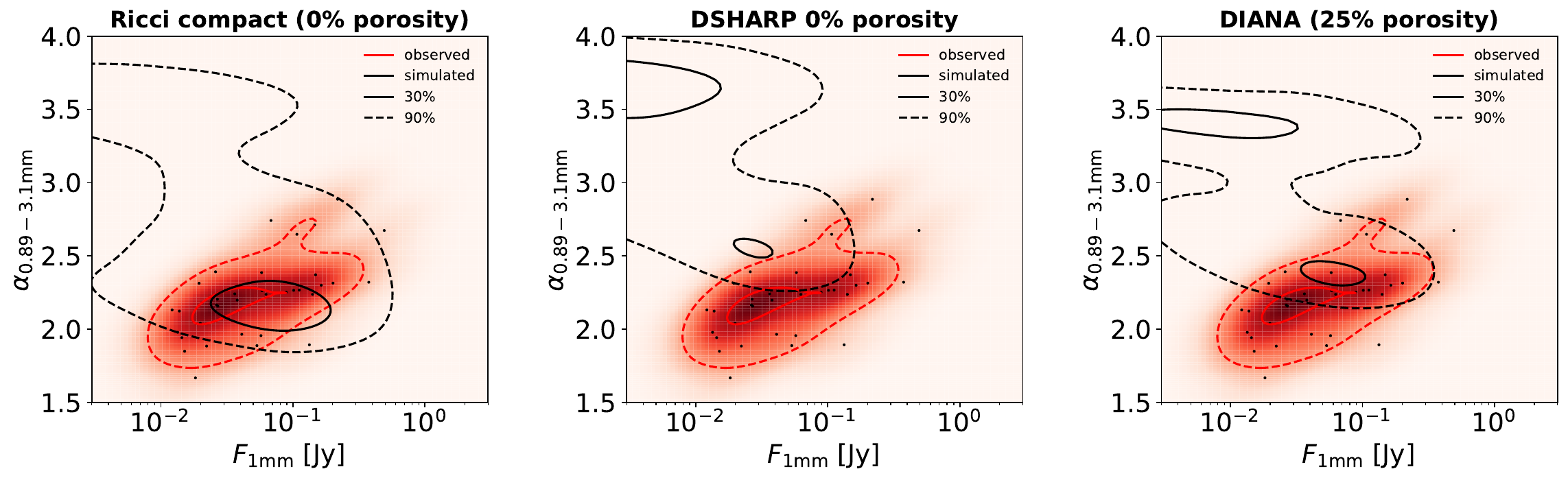}
    \caption{Spectral index distribution sub-structured disks, for the entire parameter space of initial conditions (Table \ref{table:2}) for three different opacities: Ricci opacity model \citep{ricci2010dust_a} with compact grains (Ricci compact model) as in \cite{Rosotti2019}($0\%$ grain porosity), DSHARP \citep{Birnstiel2018} with $0\%$ grain porosity and DIANA \citep{woitke2016consistent}($25\%$ grain porosity) opacities. Heatmap of the observed disks with the black dots representing each observed disk. The black and red lines refer to the simulated results and the observational results respectively. In particular, the continuous lines encompass the $30\%$ of the cumulative sum of the disks produced from the simulations or observed. The dashed lines encompass the $90\%$ instead.}
    \label{spectral_3opacity}
\end{figure*}
\subsubsection{Opacities}
All the results shown in the previous section were obtained considering Ricci compact opacities (\cite{Rosotti2019}). This opacity proved to be the best for reproducing the observed distributions, as already noted in \citet{zormpas2022large} and \citet{stadler2022impact}, for the study of the size-luminosity distribution. Figure \ref{spectral_3opacity} shows the comparison between the distributions obtained for the spectral index in the case of three different opacities: Ricci opacity model \citep{ricci2010dust_a} with compact grains (Ricci compact model) as in \cite{Rosotti2019}($0\%$ grain porosity), DSHARP \citep{Birnstiel2018} with $0\%$ grain porosity and DIANA \citep{woitke2016consistent}($25\%$ grain porosity) opacities. The only model capable of producing disks with a spectral index $\sim2.2$ is the Ricci compact model. DIANA produces disks with low spectral index values but fails to populate the observed regions at $\alpha_{0.89-3.1 mm}\leq2.3$. DSHARP suffers from the same problem as DIANA and also produces disks with a lower flux that do not reach the observed extent. The latter problem was already highlighted by \citet{zormpas2022large}, who also showed that as grain opacity increases, disk flux reduces for the DSHARP case (see appendix \ref{appendix:dsharp_porosities} for an in-depth investigation of different \% of grain porosity for DSHARP opacities). The difference in the final result between DIANA and Ricci compact lies in the compactness of the grains considered for Ricci. Instead, the difference between Ricci and DSHARP resides in the fact that Ricci's opacity has a value $\sim8.5$ higher than DHSARP's at the position of the opacity cliff if, for example, we consider a wavelength of $850\mu m$. This explains the difference in the final flux value of the disks obtained in the case of Ricci compared to DSHARP.\\
\\
Disk population synthesis represents a valuable tool that can provide additional insight into the definition of opacity and dust composition models. Dust composition, porosity and opacities determinations represent one of the main open problems in the field. Indeed, they are crucial hypotheses for determining disk characteristics, but we still lack precise knowledge of both of them. Thus, the solution to this problem constitutes a very active field in the protoplanetary disk panorama, which is being addressed through different techniques such as: sub-mm polarisation \citep{kataoka2016submillimeter,kataoka2017evidence}, scattered light phase function \citep{ginski2023observed} and multi-wavelength studies \citep{guidi2022distribution}. The multi-wavelength study by \cite{guidi2022distribution} and the scattered light phase function study by \cite{ginski2023observed} support the idea of low-porosity grains in protoplanetary disks. In particular, \cite{guidi2022distribution} study on grain properties in the ringed disk of HD 163296 shows that low porosity grains better reproduce the observations of HD 163296. \cite{ginski2023observed} shows that two categories of aggregates can be associated with polarized phase functions. The first category consists of fractal, porous aggregates, while the second consists of more compact and less porous aggregates. In particular, they note that disks belonging to the second category host embedded planets which may trigger enhanced vertical mixing leading to the production of more compact particles. Instead, sub-mm polarisation studies \citep{kataoka2016submillimeter,kataoka2017evidence} provide important information about the maximum grain size. They constrain a maximum grain size $\sim 100\mu m$. In an optically thin regime, such grains would not produce a low spectral index value; indeed $\alpha_{0.89-3.1 mm}\leq2.5$ requires grains larger than $1mm$. However, as Fig. \ref{opticaldepth} shows, low spectral index values are produced by the presence of an optically thick region caused by substructure.
\subsubsection{IMF}
\begin{figure*}
    \sidecaption
    \includegraphics[width=12cm]{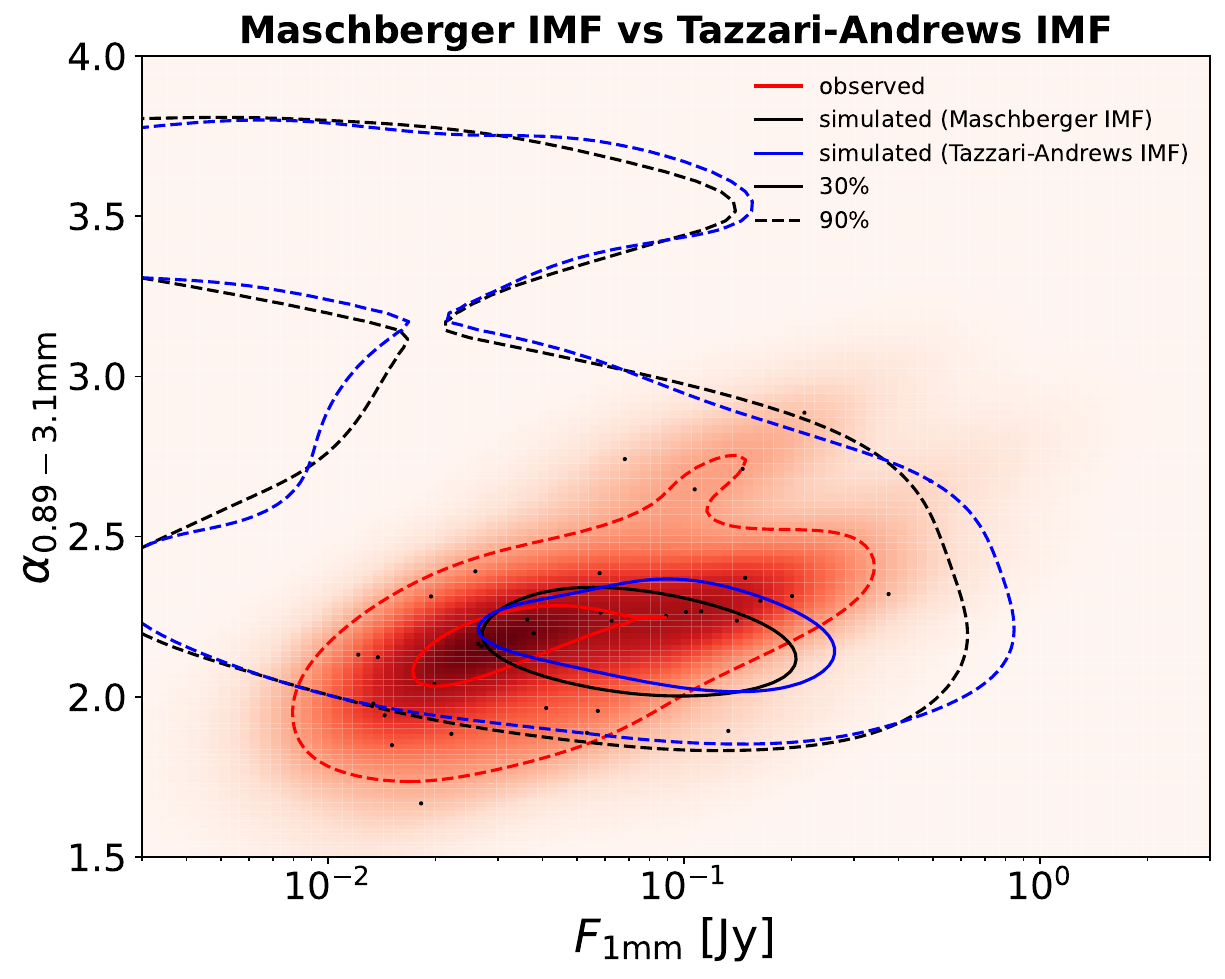}
    \caption{Spectral index distribution sub-structured disks, for the entire parameter space of initial conditions (Table \ref{table:2}) for two different IMF. Heatmap of the observed disks with the black dots representing each single observed disk. The black, blue and red lines refer to the simulated results obtained for the functional form of the IMF proposed by \cite{maschberger2013function} (black) which is based on the standard IMFs of Kroupa (\cite{kroupa2001variation},\cite{kroupa2002initial}) and Chabrier \citep{chabrier2003galactic}, the Andrews-Tazzari IMF (blue) and the observational results (red). In particular, the continuous lines encompass the $30\%$ of the cumulative sum of the disks produced from the simulations or observed. The dashed lines encompass the $90\%$ instead.}
    \label{imf_comparison}
\end{figure*}

As shown in Table \ref{table:2}, the stellar masses of our simulations were drawn from the functional form of the IMF proposed by \cite{maschberger2013function} which is based on the standard IMFs of Kroupa (\cite{kroupa2001variation},\cite{kroupa2002initial}) and Chabrier \citep{chabrier2003galactic}. To make our comparison to the observed disk distributions more consistent, we also accounted for the fact that the observed disks may be characterized by a stellar mass distribution that differs from the standard IMFs of Kroupa and Chabrier. We therefore constructed an IMF from a Kernel Density Estimate derived from the stellar mass distributions of the samples in \citet{Tazzari2021} and \citet{Andrews2018a}. Figure \ref{imf_comparison} shows that the results obtained with the new IMF deviate only slightly from the classical IMFs case, mainly by a slight shift to higher fluxes.
\subsubsection{Double substructure}
\begin{figure*}
    \sidecaption
    \includegraphics[width=12cm]{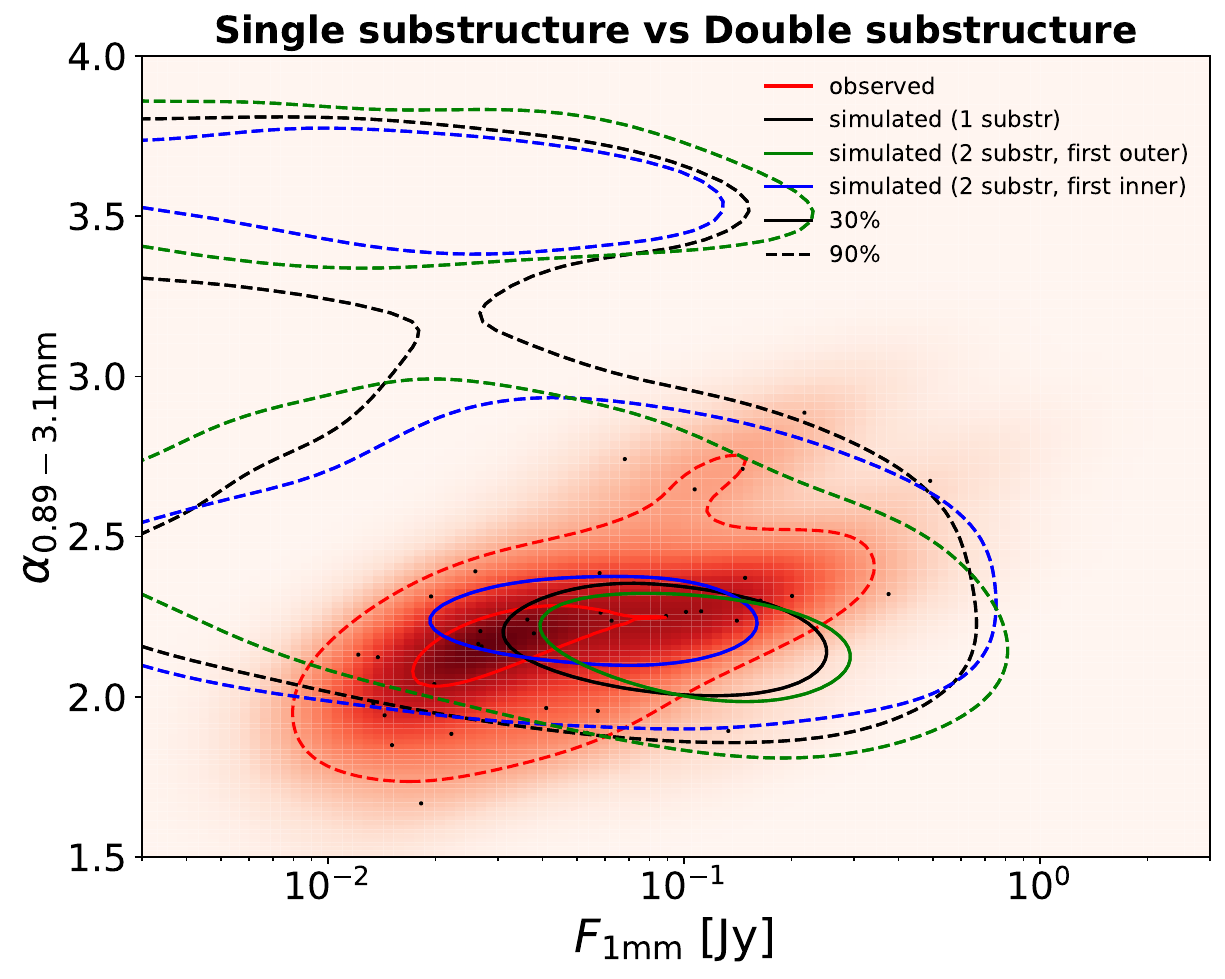}
    \caption{Spectral index distribution sub-structured disks, for the entire parameter space of initial conditions (Table \ref{table:2}) for disks with one substructure and two substructures. Heatmap of the observed disks with the black dots representing each single observed disk. The black, blue, green and red lines refer to the simulated results obtained for single substructured disks (black), simulated results obtained for double substructured disks with the inner planet inserted first (blue), double substructured disks with the outer planet inserted first (green) and the observational results (red). In particular, the continuous lines encompass the $30\%$ of the cumulative sum of the disks produced from the simulations or observed. The dashed lines encompass the $90\%$ instead.}
    \label{substr_comparison}
\end{figure*}

Figure \ref{substr_comparison} shows the comparison of the behavior of the simulated spectral index distribution when one or two planets (i.e., substructures) are inserted during the evolution of the disk. We have analyzed two different scenarios: first scenario the earliest substructure is the innermost (blue contours), second scenario the earliest substructure is the outermost (green contours). It can be immediately observed that it is possible to produce spectral index values in the observed region even in the generic case of two substructures. In general, the double substructure distributions are similar to the single substructure case and simply show a shift in the final flux produced. In the case of two substructures and faster insertion of the innermost substructure, the value of the final flux produced is lower than in the case of faster insertion of the outer substructure, as in the latter case there is a greater accumulation of material in the outer region with which a greater area is associated. A brighter disk is therefore generally obtained in the second scenario. The double substructure case with primary insertion of the innermost substructure, produces generally fainter disks than the single substructure case because a range of $r_{p}=0.05-0.5\ r_{c}$ was selected for the double substructure case.
\subsection{Future perspectives and open problems: spectral index at longer wavelength, disk size and MHD disk winds} \label{future_subsection}
\begin{figure*}
    \sidecaption
     \includegraphics[width=12cm]{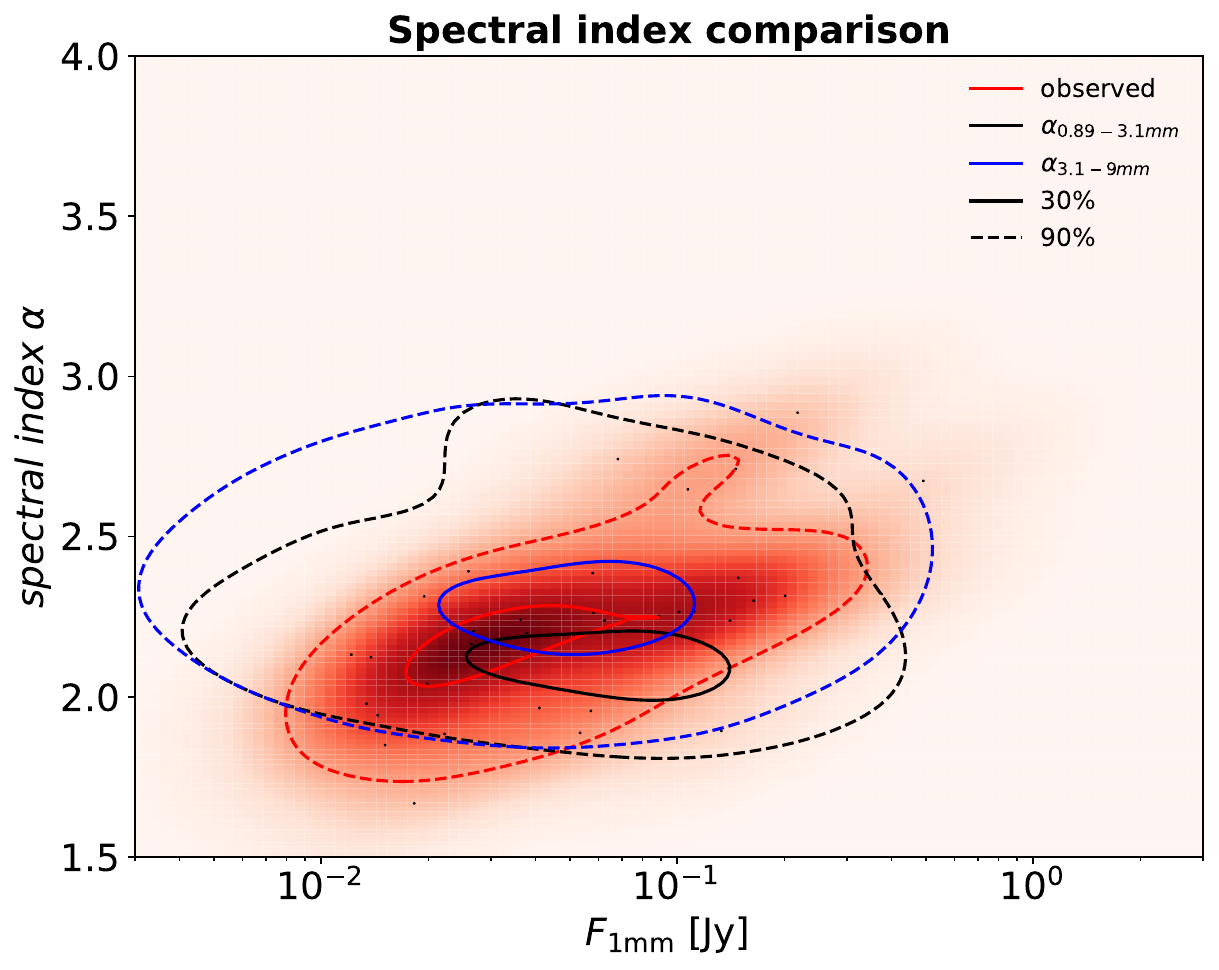}
     \caption{Spectral index distribution sub-structured disks, selecting disks with: $10^{-3.5}\leqslant\alpha\leqslant 10^{-2.5}$, $10^{-2.3}M_{\star}\leqslant M_{disk}\leqslant10^{-0.5}M_{\star}$, $v_{frag}\geq500 cm/s$, $m_{p}\geq150M_{\oplus}$, $r_{p}\leq0.75\rc$. Heatmap of the observed disks with the black dots representing each single observed disk. The black, blue and red lines refer to the simulated results obtained for $\alpha_{0.89-3.1mm}$ (black), simulated results obtained for $\alpha_{3.1-9mm}$ (blue) and the observational results ($\alpha_{0.89-3.1mm}$) (red). In particular, the continuous lines encompass the $30\%$ of the cumulative sum of the disks produced from the simulations or observed. The dashed lines encompass the $90\%$ instead.}
     \label{longer_lambda}
\end{figure*}

\begin{figure*}
    \centering
    \includegraphics[scale=0.45]{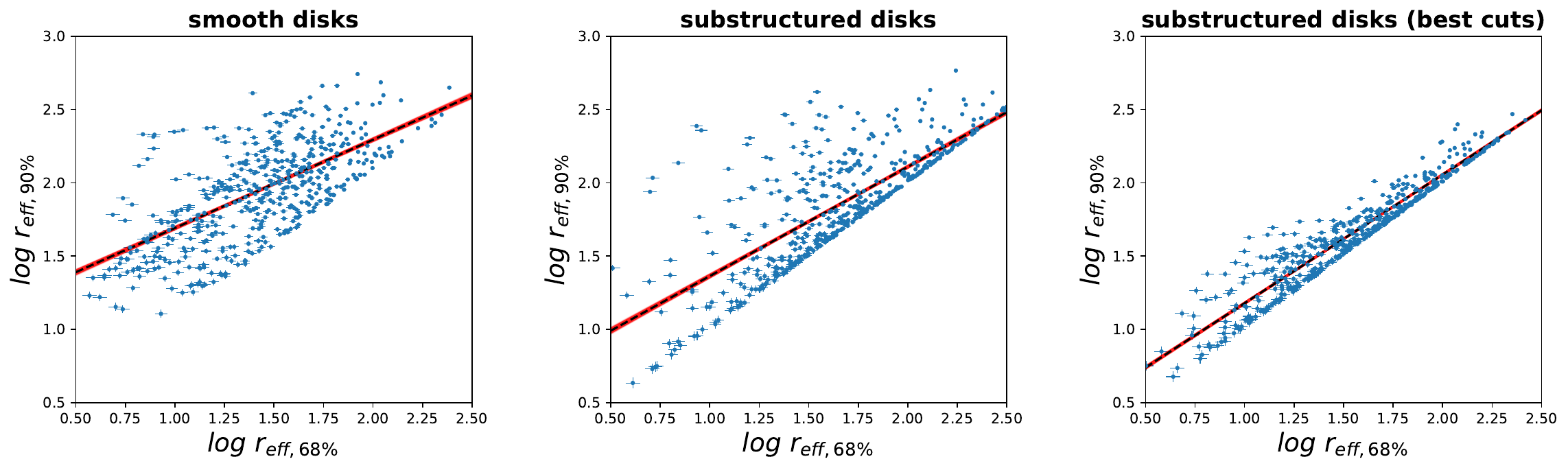}
    \caption{Comparison between $log r_{eff,90\%}$ and $log r_{eff,68\%}$ dust disk sizes for smooth disks (left panel), entire set of sub-structured disks (middle panel), sub-structured disks selected from our best case (i.e., $10^{-3.5}\leqslant\alpha\leqslant 10^{-2.5}$, $10^{-2.3}M_{\star}\leqslant M_{disk}\leqslant10^{-0.5}M_{\star}$, $v_{frag}\geq500 cm/s$, $m_{p}\geq150M_{\oplus}$, $r_{p}\leq0.75\rc$)(right plot). Blue dots represent a subset of the disks used for the estimate of the correlation between $r_{eff,90\%}$ and $r_{eff,68\%}$. The best fit obtained exploiting linmix implementation of the Bayesian linear regression
    method developed by \cite{kelly2007some} is shown as a black line, with a $1\sigma$ confidence interval reported in red.}
    \label{diskradii_comparison}
\end{figure*}

In Fig. \ref{longer_lambda} we investigate the distribution of the spectral index evaluated at longer wavelengths ($\alpha_{3.1-9mm}$) for the disk filtered by applying the best cuts introduced in Subsect. \ref{spectral_index}. We observe a slight general shift towards larger values of the spectral index since less of the emitting region produced by the substructure is optically thick and some parts become optically thin while not being made of very large grains. This behavior opens an interesting window towards disk mass estimates since a search for optically thin emission is required for accurate estimate of disk masses, and thus also for the estimate of the amount of material available for planet formation. However, at the moment, only a small sample of disks have been observed at large wavelengths, for instance only around 30 disks have been observed at $\lambda \sim 7.5mm$ (mostly by \cite{rodmann2006large} and \cite{ubach2012grain}).
\\
\\
In this study we have chosen to define our effective radius for defining disk size as the radius that encloses the $68\%$ of the total amount of flux produced by the disk ($r_{eff,68\%}$), following \citet{Tripathi2017}. Nevertheless, most recent observations have considered the radius that encloses $90\%$ of the emission ($r_{eff,90\%}$). However, \cite{hendler2020evolution} finds a 1-1 correlation between $r_{eff,90\%}$ and $r_{eff,68\%}$(see Fig. 15 in \citep{hendler2020evolution}). We have thus evaluated $r_{eff,90\%}$ for each disk and tested if our population synthesis can reproduce this observed trend. Figure \ref{diskradii_comparison} shows the behavior exhibited by three different disk populations: smooth disks, substructured disks without applying any cut to the initial conditions and substructured disks filtered with our best conditions (i.e., $10^{-3.5}\leqslant\alpha\leqslant 10^{-2.5}$, $10^{-2.3}M_{\star}\leqslant M_{disk}\leqslant10^{-0.5}M_{\star}$, $v_{frag}\geq500 cm/s$, $m_{p}\geq150M_{\oplus}$, $r_{p}\leq0.75\rc$). We select a set of $10^{4}$ disk per population and fit each sample exploiting linmix implementation of the Bayesian linear regression method developed by \cite{kelly2007some}. Results reported in Table \ref{table:3} show that smooth disks do not reproduce the correlation observed in \cite{hendler2020evolution}, same applies to the sub-structured disks population. However, filtering the sub-structured disk population selecting disks with $10^{-3.5}\leqslant\alpha\leqslant 10^{-2.5}$, $10^{-2.3}M_{\star}\leqslant M_{disk}\leqslant10^{-0.5}M_{\star}$, $v_{frag}\geq500 cm/s$, $m_{p}\geq150M_{\oplus}$, $r_{p}\leq0.75\rc$, that is applying the best cuts introduced in Subsect. \ref{spectral_index} which lead to a matching to both the spectral index and size-luminosity distribution, we obtain a correlation between $log r_{eff,90\%}$ and $log r_{eff,68\%}$. This result further strengthens the outcomes outlined in Subsect. \ref{spectral_index} concerning the need for substructure in protoplanetary disks and the associated parameters space.
\begin{table*}  
\caption{$log r_{eff,90\%}$ vs $log r_{eff,68\%}$ fit results.}  
\label{table:3}      
\centering                          
\begin{tabular}{c c c c c}        
\hline\hline                 
Disk population & intercept & slope & regression & linear \\
& & & intrinsic scatter & correlation coefficient\\    
\hline                        
smooth disks & 1.09$\pm$0.01 & 0.60$\pm$0.01 & 0.06$\pm$0.00 & 0.67$\pm$0.01\\
sub-structured disks & 0.62$\pm$0.01 & 0.75$\pm$0.01 & 0.06$\pm$0.00 & 0.78$\pm$0.00\\
sub-structured disks (best cuts) & 0.30$\pm$0.00 & 0.88$\pm$0.00 & 0.01$\pm$0.00 & 0.95$\pm$0.00\\
\hline                                   
\end{tabular}
\tablefoot{$log r_{eff,90\%}$ vs $log r_{eff,68\%}$ fit results for three different disk populations: smooth disks, substructured disks without applying any cut to the initial conditions and substructured disks filtered with our best cuts (i.e., $10^{-3.5}\leqslant\alpha\leqslant 10^{-2.5}$, $10^{-2.3}M_{\star}\leqslant M_{disk}\leqslant10^{-0.5}M_{\star}$, $v_{frag}\geq500 cm/s$, $m_{p}\geq150M_{\oplus}$, $r_{p}\leq0.75\rc$). Fit results obtained exploiting linmix implementation of the Bayesian linear regression method developed by \cite{kelly2007some}.}
\end{table*}
\\
\\
In this work we focused on the classical scenario in which disks evolve viscously. However, in recent years, the hypothesis that the evolution of the disk is driven by magnetic winds has become increasingly popular. We therefore aim to expand our investigation on the MHD disk winds scenario in the future, to determine if and what differences might arise compared to the viscous scenario. In this respect, \cite{zagaria2022modelling} shows that current available observations do not allow discerning between viscous and magnetic wind scenarios and that from the dust perspective, there is little difference between the viscous case and MHD winds. Indeed, they show that SLR can be reproduced even by MHD disk winds models, except for the very large disks, which can, however, be explained assuming the presence of substructures. We therefore expect, adopting an MHD wind model, similar conclusions to those obtained for the viscous scenario.
\clearpage

\section{Conclusions} \label{conclusions}
In this work, we conducted a study aimed at understanding the possibility of reproducing the observed spectral index distribution of protoplanetary disks. We exploited the two-pop-py 1D evolutionary model for the dust and gas in protoplanetary disks. We considered both smooth and substructured disks and a wide initial parameter space. We firstly compared the simulated distribution obtained for the spectral index to the observed ones reported in \citet{tazzari2021first} and then analysed the possibility of matching also the size-luminosity distribution considering the observed distribution reported in \citet{Andrews2018a}. We have been able to identify the initial conditions and the kind of disks needed to match the spectral index distribution and in particular to match simultaneously also the size-luminosity distribution. These are the main results we have outlined:
\begin{enumerate}
    \item Substructures are needed to produce small values of the spectral index in the range of the observed ones; smooth disks produce only large values of the spectral index (Fig. \ref{smooth_vs_substr_earlylate}).
    \item The substructure has to be formed quickly, that is within $\sim 0.4Myr$ (Fig. \ref{smooth_vs_substr_earlylate}) to produce a value of the spectral index below 2.5.
    \item Filtering the substructure disks with $10^{-3.5}\leqslant\alpha\leqslant 10^{-2.5}$, $10^{-2.3}M_{\star}\leqslant M_{disk}\leqslant10^{-0.5}M_{\star}$, $v_{frag}\geq500 cm/s$, $m_{p}\geq150M_{\oplus}$, $r_{p}\leq0.75\rc$ we obtain a match between the spectral index simulated distribution and the observed distribution (Fig. \ref{spectral_finalcut}), proving that it is possible to reproduce the observed distribution for a reasonable range of initial conditions.
    \item An in-depth investigation of the real cause behind the production of low spectral index values for substructured disks revealed that this is achieved by the production of an optically thick region in the disk, originating from the accumulation of material due to the presence of substructure.
    \item The matching obtained between the simulated and observed spectral index distribution automatically ensures a matching between the corresponding simulated and observed size-luminosity distribution.
    \item It is possible to reproduce the size-luminosity distribution with a population of only substructured disks, thus, no longer requiring the mix of smooth and substructured disks proposed in \cite{zormpas2022large}. 
    \item The 1-1 correlation between the $r_{eff,90\%}$ and $r_{eff,68\%}$ observed in \cite{hendler2020evolution} cannot be reproduced by smooth disks and by the entire sample of sub-structured disks, but it can be retrieved filtering the sub-structured disk sample with the same parameter ranges that lead to reproducing both the spectral index and size-luminosity distributions. 
    \item Studying different opacities (Ricci compact \citealt{Rosotti2019},DSHARP \citealt{Birnstiel2018},DIANA \citealt{woitke2016consistent}) we showed that the only one capable of leading to a matching of the spectral index distribution is the Ricci compact opacity. Only opacities with high absorption efficiency can reproduce the observed spectral indices.
    \item Disks with two substructures can match the spectral index distribution, showing a behavior similar to the single substructure case. 
\end{enumerate}
This study shows that it is possible to reproduce the observed distributions for both spectral index and size-luminosity, extending the results obtained for individual disk studies to the broader level of a disk population synthesis. 

\begin{acknowledgements}
L.D. and T.B. acknowledge funding by the Deutsche Forschungsgemeinschaft (DFG, German Research Foundation) under grant 325594231 and Germany's Excellence Strategy - EXC-2094 - 390783311. T.B. acknowledges funding from the European Research Council (ERC) under the European Union's Horizon 2020 research and innovation programme under grant agreement No 714769. \\
PP acknowledges funding from the UK Research and Innovation (UKRI) under the UK government’s Horizon Europe funding guarantee from ERC (under grant agreement No 101076489).\\
GR acknowledges funding from the Fondazione Cariplo, grant no. 2022-1217, and the European Research Council (ERC) under the European Union’s Horizon Europe Research \& Innovation Programme under grant agreement no. 101039651 (DiscEvol). Views and opinions expressed are however those of the author(s) only, and do not necessarily reflect those of the European Union or the European Research Council Executive Agency. Neither the European Union nor the granting authority can be held responsible for them.
\end{acknowledgements}

%
%

\bibliographystyle{aa}
\bibliography{biblio}

\begin{thebibliography}{70}
\expandafter\ifx\csname natexlab\endcsname\relax\def\natexlab#1{#1}\fi

\bibitem[{{Andrews} {et~al.}(2018{\natexlab{a}}){Andrews}, {Huang}, {P{\'e}rez}, {Isella}, {Dullemond}, {Kurtovic}, {Guzm{\'a}n}, {Carpenter}, {Wilner}, {Zhang}, {Zhu}, {Birnstiel}, {Bai}, {Benisty}, {Hughes}, {{\"O}berg}, \& {Ricci}}]{Andrews2018}
{Andrews}, S.~M., {Huang}, J., {P{\'e}rez}, L.~M., {et~al.} 2018{\natexlab{a}}, \apjl, 869, L41

\bibitem[{Andrews {et~al.}(2013)Andrews, Rosenfeld, Kraus, \& Wilner}]{andrews2013mass}
Andrews, S.~M., Rosenfeld, K.~A., Kraus, A.~L., \& Wilner, D.~J. 2013, The Astrophysical Journal, 771, 129

\bibitem[{{Andrews} {et~al.}(2018{\natexlab{b}}){Andrews}, {Terrell}, {Tripathi}, {Ansdell}, {Williams}, \& {Wilner}}]{Andrews2018a}
{Andrews}, S.~M., {Terrell}, M., {Tripathi}, A., {et~al.} 2018{\natexlab{b}}, \apj, 865, 157

\bibitem[{Ansdell {et~al.}(2016)Ansdell, Williams, van~der Marel, Carpenter, Guidi, Hogerheijde, Mathews, Manara, Miotello, Natta, {et~al.}}]{ansdell2016alma}
Ansdell, M., Williams, J.~P., van~der Marel, N., {et~al.} 2016, The Astrophysical Journal, 828, 46

\bibitem[{{Bai} \& {Stone}(2014)}]{Bai2014}
{Bai}, X.-N. \& {Stone}, J.~M. 2014, \apj, 796, 31

\bibitem[{Birnstiel {et~al.}(2015)Birnstiel, Andrews, Pinilla, \& Kama}]{birnstiel2015dust}
Birnstiel, T., Andrews, S.~M., Pinilla, P., \& Kama, M. 2015, The Astrophysical Journal Letters, 813, L14

\bibitem[{{Birnstiel} {et~al.}(2010){Birnstiel}, {Dullemond}, \& {Brauer}}]{Birnstiel2010}
{Birnstiel}, T., {Dullemond}, C.~P., \& {Brauer}, F. 2010, \aap, 513, A79

\bibitem[{{Birnstiel} {et~al.}(2018){Birnstiel}, {Dullemond}, {Zhu}, {Andrews}, {Bai}, {Wilner}, {Carpenter}, {Huang}, {Isella}, {Benisty}, {P{\'e}rez}, \& {Zhang}}]{Birnstiel2018}
{Birnstiel}, T., {Dullemond}, C.~P., {Zhu}, Z., {et~al.} 2018, \apjl, 869, L45

\bibitem[{Birnstiel {et~al.}(2018)Birnstiel, Dullemond, Zhu, Andrews, Bai, Wilner, Carpenter, Huang, Isella, Benisty, {et~al.}}]{birnstiel2018disk}
Birnstiel, T., Dullemond, C.~P., Zhu, Z., {et~al.} 2018, The Astrophysical Journal Letters, 869, L45

\bibitem[{Birnstiel {et~al.}(2012)Birnstiel, Klahr, \& Ercolano}]{birnstiel2012simple}
Birnstiel, T., Klahr, H., \& Ercolano, B. 2012, Astronomy \& Astrophysics, 539, A148

\bibitem[{Chabrier(2003)}]{chabrier2003galactic}
Chabrier, G. 2003, Publications of the Astronomical Society of the Pacific, 115, 763

\bibitem[{{D'Alessio} {et~al.}(2006){D'Alessio}, {Calvet}, {Hartmann}, {Franco-Hern{\'a}ndez}, \& {Serv{\'\i}n}}]{Dalessio2006}
{D'Alessio}, P., {Calvet}, N., {Hartmann}, L., {Franco-Hern{\'a}ndez}, R., \& {Serv{\'\i}n}, H. 2006, \apj, 638, 314

\bibitem[{Draine(2006)}]{draine2006submillimeter}
Draine, B. 2006, The Astrophysical Journal, 636, 1114

\bibitem[{Ginski {et~al.}(2023)Ginski, Tazaki, Dominik, \& Stolker}]{ginski2023observed}
Ginski, C., Tazaki, R., Dominik, C., \& Stolker, T. 2023, The Astrophysical Journal, 953, 92

\bibitem[{Guidi {et~al.}(2022)Guidi, Isella, Testi, Chandler, Liu, Schmid, Rosotti, Meng, Jennings, Williams, {et~al.}}]{guidi2022distribution}
Guidi, G., Isella, A., Testi, L., {et~al.} 2022, Astronomy \& Astrophysics, 664, A137

\bibitem[{Hendler {et~al.}(2020)Hendler, Pascucci, Pinilla, Tazzari, Carpenter, Malhotra, \& Testi}]{hendler2020evolution}
Hendler, N., Pascucci, I., Pinilla, P., {et~al.} 2020, The Astrophysical Journal, 895, 126

\bibitem[{{Huang} {et~al.}(2018){Huang}, {Andrews}, {Dullemond}, {Isella}, {P{\'e}rez}, {Guzm{\'a}n}, {{\"O}berg}, {Zhu}, {Zhang}, {Bai}, {Benisty}, {Birnstiel}, {Carpenter}, {Hughes}, {Ricci}, {Weaver}, \& {Wilner}}]{Huang2018}
{Huang}, J., {Andrews}, S.~M., {Dullemond}, C.~P., {et~al.} 2018, \apjl, 869, L42

\bibitem[{Izquierdo {et~al.}(2022)Izquierdo, Facchini, Rosotti, van Dishoeck, \& Testi}]{izquierdo2022new}
Izquierdo, A.~F., Facchini, S., Rosotti, G.~P., van Dishoeck, E.~F., \& Testi, L. 2022, The Astrophysical Journal, 928, 2

\bibitem[{{Johansen} {et~al.}(2009){Johansen}, {Youdin}, \& {Klahr}}]{Johansen2009}
{Johansen}, A., {Youdin}, A., \& {Klahr}, H. 2009, \apj, 697, 1269

\bibitem[{Kanagawa {et~al.}(2016)Kanagawa, Muto, Tanaka, Tanigawa, Takeuchi, Tsukagoshi, \& Momose}]{kanagawa2016mass}
Kanagawa, K.~D., Muto, T., Tanaka, H., {et~al.} 2016, Publications of the Astronomical Society of Japan, 68, 43

\bibitem[{Kataoka {et~al.}(2015)Kataoka, Muto, Momose, Tsukagoshi, Fukagawa, Shibai, Hanawa, Murakawa, \& Dullemond}]{kataoka2015millimeter}
Kataoka, A., Muto, T., Momose, M., {et~al.} 2015, The Astrophysical Journal, 809, 78

\bibitem[{Kataoka {et~al.}(2016)Kataoka, Tsukagoshi, Momose, Nagai, Muto, Dullemond, Pohl, Fukagawa, Shibai, Hanawa, {et~al.}}]{kataoka2016submillimeter}
Kataoka, A., Tsukagoshi, T., Momose, M., {et~al.} 2016, The Astrophysical Journal Letters, 831, L12

\bibitem[{Kataoka {et~al.}(2017)Kataoka, Tsukagoshi, Pohl, Muto, Nagai, Stephens, Tomisaka, \& Momose}]{kataoka2017evidence}
Kataoka, A., Tsukagoshi, T., Pohl, A., {et~al.} 2017, The Astrophysical Journal Letters, 844, L5

\bibitem[{Kelly(2007)}]{kelly2007some}
Kelly, B.~C. 2007, The Astrophysical Journal, 665, 1489

\bibitem[{Kenyon {et~al.}(1996)Kenyon, Yi, \& Hartmann}]{kenyon1996magnetic}
Kenyon, S.~J., Yi, I., \& Hartmann, L. 1996, The Astrophysical Journal, 462, 439

\bibitem[{{Keppler} {et~al.}(2018){Keppler}, {Benisty}, {M{\"u}ller}, {Henning}, {van Boekel}, {Cantalloube}, {Ginski}, {van Holstein}, {Maire}, {Pohl}, {Samland}, {Avenhaus}, {Baudino}, {Boccaletti}, {de Boer}, {Bonnefoy}, {Chauvin}, {Desidera}, {Langlois}, {Lazzoni}, {Marleau}, {Mordasini}, {Pawellek}, {Stolker}, {Vigan}, {Zurlo}, {Birnstiel}, {Brandner}, {Feldt}, {Flock}, {Girard}, {Gratton}, {Hagelberg}, {Isella}, {Janson}, {Juhasz}, {Kemmer}, {Kral}, {Lagrange}, {Launhardt}, {Matter}, {M{\'e}nard}, {Milli}, {Molli{\`e}re}, {Olofsson}, {P{\'e}rez}, {Pinilla}, {Pinte}, {Quanz}, {Schmidt}, {Udry}, {Wahhaj}, {Williams}, {Buenzli}, {Cudel}, {Dominik}, {Galicher}, {Kasper}, {Lannier}, {Mesa}, {Mouillet}, {Peretti}, {Perrot}, {Salter}, {Sissa}, {Wildi}, {Abe}, {Antichi}, {Augereau}, {Baruffolo}, {Baudoz}, {Bazzon}, {Beuzit}, {Blanchard}, {Brems}, {Buey}, {De Caprio}, {Carbillet}, {Carle}, {Cascone}, {Cheetham}, {Claudi}, {Costille}, {Delboulb{\'e}}, {Dohlen}, {Fantinel}, {Feautrier}, {Fusco}, {Giro}, {Gluck},
  {Gry}, {Hubin}, {Hugot}, {Jaquet}, {Le Mignant}, {Llored}, {Madec}, {Magnard}, {Martinez}, {Maurel}, {Meyer}, {M{\"o}ller-Nilsson}, {Moulin}, {Mugnier}, {Orign{\'e}}, {Pavlov}, {Perret}, {Petit}, {Pragt}, {Puget}, {Rabou}, {Ramos}, {Rigal}, {Rochat}, {Roelfsema}, {Rousset}, {Roux}, {Salasnich}, {Sauvage}, {Sevin}, {Soenke}, {Stadler}, {Suarez}, {Turatto}, \& {Weber}}]{Keppler2018}
{Keppler}, M., {Benisty}, M., {M{\"u}ller}, A., {et~al.} 2018, \aap, 617, A44

\bibitem[{Kroupa(2001)}]{kroupa2001variation}
Kroupa, P. 2001, Monthly Notices of the Royal Astronomical Society, 322, 231

\bibitem[{Kroupa(2002)}]{kroupa2002initial}
Kroupa, P. 2002, Science, 295, 82

\bibitem[{{Long} {et~al.}(2018){Long}, {Pinilla}, {Herczeg}, {Harsono}, {Dipierro}, {Pascucci}, {Hendler}, {Tazzari}, {Ragusa}, {Salyk}, {Edwards}, {Lodato}, {van de Plas}, {Johnstone}, {Liu}, {Boehler}, {Cabrit}, {Manara}, {Menard}, {Mulders}, {Nisini}, {Fischer}, {Rigliaco}, {Banzatti}, {Avenhaus}, \& {Gully-Santiago}}]{Long2018}
{Long}, F., {Pinilla}, P., {Herczeg}, G.~J., {et~al.} 2018, \apj, 869, 17

\bibitem[{{L{\"u}st}(1952)}]{Lust1952}
{L{\"u}st}, R. 1952, Zeitschrift Naturforschung Teil A, 7, 87

\bibitem[{Lynden-Bell \& Pringle(1974)}]{lynden1974evolution}
Lynden-Bell, D. \& Pringle, J.~E. 1974, Monthly Notices of the Royal Astronomical Society, 168, 603

\bibitem[{{Lynden-Bell} \& {Pringle}(1974)}]{Lynden-Bell1974}
{Lynden-Bell}, D. \& {Pringle}, J.~E. 1974, \mnras, 168, 603

\bibitem[{Manara {et~al.}(2016)Manara, Rosotti, Testi, Natta, Alcal{\'a}, Williams, Ansdell, Miotello, Van Der~Marel, Tazzari, {et~al.}}]{manara2016evidence}
Manara, C., Rosotti, G., Testi, L., {et~al.} 2016, Astronomy \& Astrophysics, 591, L3

\bibitem[{{Manara} {et~al.}(2023){Manara}, {Ansdell}, {Rosotti}, {Hughes}, {Armitage}, {Lodato}, \& {Williams}}]{Manara2023}
{Manara}, C.~F., {Ansdell}, M., {Rosotti}, G.~P., {et~al.} 2023, in Astronomical Society of the Pacific Conference Series, Vol. 534, Protostars and Planets VII, ed. S.~{Inutsuka}, Y.~{Aikawa}, T.~{Muto}, K.~{Tomida}, \& M.~{Tamura}, 539

\bibitem[{Maschberger(2013)}]{maschberger2013function}
Maschberger, T. 2013, Monthly Notices of the Royal Astronomical Society, 429, 1725

\bibitem[{Miotello {et~al.}(2023)Miotello, Kamp, Birnstiel, Cleeves, \& Kataoka}]{miotello2023setting}
Miotello, A., Kamp, I., Birnstiel, T., Cleeves, L., \& Kataoka, A. 2023, in Astronomical Society of the Pacific Conference Series, Vol. 534, 501

\bibitem[{Miyake \& Nakagawa(1993)}]{miyake1993effects}
Miyake, K. \& Nakagawa, Y. 1993, icarus, 106, 20

\bibitem[{{M{\"u}ller} {et~al.}(2018){M{\"u}ller}, {Keppler}, {Henning}, {Samland}, {Chauvin}, {Beust}, {Maire}, {Molaverdikhani}, {van Boekel}, {Benisty}, {Boccaletti}, {Bonnefoy}, {Cantalloube}, {Charnay}, {Baudino}, {Gennaro}, {Long}, {Cheetham}, {Desidera}, {Feldt}, {Fusco}, {Girard}, {Gratton}, {Hagelberg}, {Janson}, {Lagrange}, {Langlois}, {Lazzoni}, {Ligi}, {M{\'e}nard}, {Mesa}, {Meyer}, {Molli{\`e}re}, {Mordasini}, {Moulin}, {Pavlov}, {Pawellek}, {Quanz}, {Ramos}, {Rouan}, {Sissa}, {Stadler}, {Vigan}, {Wahhaj}, {Weber}, \& {Zurlo}}]{Muller2018}
{M{\"u}ller}, A., {Keppler}, M., {Henning}, T., {et~al.} 2018, \aap, 617, L2

\bibitem[{Natta {et~al.}(2004)Natta, Testi, Johnstone, Adams, Lin, Neufeeld, \& Ostriker}]{natta2004star}
Natta, A., Testi, L., Johnstone, D., {et~al.} 2004, in ASP Conf. Proc., Vol. 323, 279

\bibitem[{{Okuzumi} {et~al.}(2016){Okuzumi}, {Momose}, {Sirono}, {Kobayashi}, \& {Tanaka}}]{Okuzumi2016}
{Okuzumi}, S., {Momose}, M., {Sirono}, S.-i., {Kobayashi}, H., \& {Tanaka}, H. 2016, \apj, 821, 82

\bibitem[{{Paardekooper} \& {Mellema}(2004)}]{Paardekooper2004}
{Paardekooper}, S.~J. \& {Mellema}, G. 2004, \aap, 425, L9

\bibitem[{Pascucci {et~al.}(2016)Pascucci, Testi, Herczeg, Long, Manara, Hendler, Mulders, Krijt, Ciesla, Henning, {et~al.}}]{pascucci2016steeper}
Pascucci, I., Testi, L., Herczeg, G.~J., {et~al.} 2016, The Astrophysical Journal, 831, 125

\bibitem[{{Pinilla} {et~al.}(2012){Pinilla}, {Benisty}, \& {Birnstiel}}]{Pinilla2012a}
{Pinilla}, P., {Benisty}, M., \& {Birnstiel}, T. 2012, \aap, 545, A81

\bibitem[{Pinilla {et~al.}(2012)Pinilla, Birnstiel, Ricci, Dullemond, Uribe, Testi, \& Natta}]{pinilla2012trapping}
Pinilla, P., Birnstiel, T., Ricci, L., {et~al.} 2012, Astronomy \& Astrophysics, 538, A114

\bibitem[{{Pinilla} {et~al.}(2017){Pinilla}, {Pohl}, {Stammler}, \& {Birnstiel}}]{Pinilla2017}
{Pinilla}, P., {Pohl}, A., {Stammler}, S.~M., \& {Birnstiel}, T. 2017, \apj, 845, 68

\bibitem[{{Pinte} {et~al.}(2018){Pinte}, {Price}, {M{\'e}nard}, {Duch{\^e}ne}, {Dent}, {Hill}, {de Gregorio-Monsalvo}, {Hales}, \& {Mentiplay}}]{Pinte2018}
{Pinte}, C., {Price}, D.~J., {M{\'e}nard}, F., {et~al.} 2018, \apjl, 860, L13

\bibitem[{{Pollack} {et~al.}(1994){Pollack}, {Hollenbach}, {Beckwith}, {Simonelli}, {Roush}, \& {Fong}}]{Pollack1994}
{Pollack}, J.~B., {Hollenbach}, D., {Beckwith}, S., {et~al.} 1994, \apj, 421, 615

\bibitem[{{Pollack} {et~al.}(1996){Pollack}, {Hubickyj}, {Bodenheimer}, {Lissauer}, {Podolak}, \& {Greenzweig}}]{pollack1996}
{Pollack}, J.~B., {Hubickyj}, O., {Bodenheimer}, P., {et~al.} 1996, Icarus, Vol. 124, Issue 1, p. 62-85

\bibitem[{{Ragusa} {et~al.}(2017){Ragusa}, {Dipierro}, {Lodato}, {Laibe}, \& {Price}}]{Ragusa2017}
{Ragusa}, E., {Dipierro}, G., {Lodato}, G., {Laibe}, G., \& {Price}, D.~J. 2017, \mnras, 464, 1449

\bibitem[{Ricci {et~al.}(2010{\natexlab{a}})Ricci, Testi, Natta, \& Brooks}]{ricci2010dust_b}
Ricci, L., Testi, L., Natta, A., \& Brooks, K. 2010{\natexlab{a}}, Astronomy \& Astrophysics, 521, A66

\bibitem[{Ricci {et~al.}(2010{\natexlab{b}})Ricci, Testi, Natta, Neri, Cabrit, \& Herczeg}]{ricci2010dust_a}
Ricci, L., Testi, L., Natta, A., {et~al.} 2010{\natexlab{b}}, Astronomy \& Astrophysics, 512, A15

\bibitem[{{Rice} {et~al.}(2006){Rice}, {Armitage}, {Wood}, \& {Lodato}}]{Rice2006}
{Rice}, W.~K.~M., {Armitage}, P.~J., {Wood}, K., \& {Lodato}, G. 2006, \mnras, 373, 1619

\bibitem[{Rodmann {et~al.}(2006)Rodmann, Henning, Chandler, Mundy, \& Wilner}]{rodmann2006large}
Rodmann, J., Henning, T., Chandler, C., Mundy, L., \& Wilner, D. 2006, Astronomy \& Astrophysics, 446, 211

\bibitem[{{Rosotti} {et~al.}(2019a){Rosotti}, {Booth}, {Tazzari}, {Clarke}, {Lodato}, \& {Testi}}]{Rosotti2019}
{Rosotti}, G.~P., {Booth}, R.~A., {Tazzari}, M., {et~al.} 2019a, \mnras, 486, L63

\bibitem[{Rosotti {et~al.}(2019b)Rosotti, Tazzari, Booth, Testi, Lodato, \& Clarke}]{rosotti2019time}
Rosotti, G.~P., Tazzari, M., Booth, R.~A., {et~al.} 2019b, Monthly Notices of the Royal Astronomical Society, 486, 4829

\bibitem[{Savvidou \& Bitsch(2023)}]{savvidou2023make}
Savvidou, S. \& Bitsch, B. 2023, Astronomy \& Astrophysics, 679, A42

\bibitem[{Shakura \& Sunyaev(1973)}]{shakura1973black}
Shakura, N.~I. \& Sunyaev, R.~A. 1973, Astronomy and Astrophysics, Vol. 24, p. 337-355, 24, 337

\bibitem[{{Shi} {et~al.}(2012){Shi}, {Krolik}, {Lubow}, \& {Hawley}}]{Shi2012}
{Shi}, J.-M., {Krolik}, J.~H., {Lubow}, S.~H., \& {Hawley}, J.~F. 2012, \apj, 749, 118

\bibitem[{Siess {et~al.}(2000)Siess, Dufour, \& Forestini}]{siess2000internet}
Siess, L., Dufour, E., \& Forestini, M. 2000, arXiv preprint astro-ph/0003477

\bibitem[{Stadler {et~al.}(2022)Stadler, G{\'a}rate, Pinilla, Lenz, Dullemond, Birnstiel, \& Stammler}]{stadler2022impact}
Stadler, J., G{\'a}rate, M., Pinilla, P., {et~al.} 2022, Astronomy \& Astrophysics, 668, A104

\bibitem[{{Tazzari} {et~al.}(2021a){Tazzari}, {Clarke}, {Testi}, {Williams}, {Facchini}, {Manara}, {Natta}, \& {Rosotti}}]{Tazzari2021}
{Tazzari}, M., {Clarke}, C.~J., {Testi}, L., {et~al.} 2021a, \mnras, 506, 2804

\bibitem[{Tazzari {et~al.}(2021b)Tazzari, Testi, Natta, Williams, Ansdell, Carpenter, Facchini, Guidi, Hogherheijde, Manara, {et~al.}}]{tazzari2021first}
Tazzari, M., Testi, L., Natta, A., {et~al.} 2021b, Monthly Notices of the Royal Astronomical Society, 506, 5117

\bibitem[{{Teague} {et~al.}(2018){Teague}, {Bae}, {Bergin}, {Birnstiel}, \& {Foreman-Mackey}}]{Teague2018}
{Teague}, R., {Bae}, J., {Bergin}, E.~A., {Birnstiel}, T., \& {Foreman-Mackey}, D. 2018, \apjl, 860, L12

\bibitem[{{Tripathi} {et~al.}(2017){Tripathi}, {Andrews}, {Birnstiel}, \& {Wilner}}]{Tripathi2017}
{Tripathi}, A., {Andrews}, S.~M., {Birnstiel}, T., \& {Wilner}, D.~J. 2017, \apj, 845, 44

\bibitem[{Ubach {et~al.}(2012)Ubach, Maddison, Wright, Wilner, Lommen, \& Koribalski}]{ubach2012grain}
Ubach, C., Maddison, S.~T., Wright, C.~M., {et~al.} 2012, Monthly Notices of the Royal Astronomical Society, 425, 3137

\bibitem[{Woitke {et~al.}(2016)Woitke, Min, Pinte, Thi, Kamp, Rab, Anthonioz, Antonellini, Baldovin-Saavedra, Carmona, {et~al.}}]{woitke2016consistent}
Woitke, P., Min, M., Pinte, C., {et~al.} 2016, Astronomy \& Astrophysics, 586, A103

\bibitem[{Youdin \& Lithwick(2007)}]{youdin2007particle}
Youdin, A.~N. \& Lithwick, Y. 2007, icarus, 192, 588

\bibitem[{Zagaria {et~al.}(2022)Zagaria, Rosotti, Clarke, \& Tabone}]{zagaria2022modelling}
Zagaria, F., Rosotti, G.~P., Clarke, C.~J., \& Tabone, B. 2022, Monthly Notices of the Royal Astronomical Society, 514, 1088

\bibitem[{Zhu {et~al.}(2012)Zhu, Nelson, Dong, Espaillat, \& Hartmann}]{zhu2012dust}
Zhu, Z., Nelson, R.~P., Dong, R., Espaillat, C., \& Hartmann, L. 2012, The Astrophysical Journal, 755, 6

\bibitem[{Zormpas {et~al.}(2022)Zormpas, Birnstiel, Rosotti, \& Andrews}]{zormpas2022large}
Zormpas, A., Birnstiel, T., Rosotti, G.~P., \& Andrews, S.~M. 2022, Astronomy \& Astrophysics, 661, A66

\end{thebibliography}

%
%


\begin{appendix}
\section{Evolving $L_{star}$ vs fixed $L_{star}$}\label{appendix:l_star}
\begin{figure*}
    \sidecaption
    \includegraphics[width=12cm]{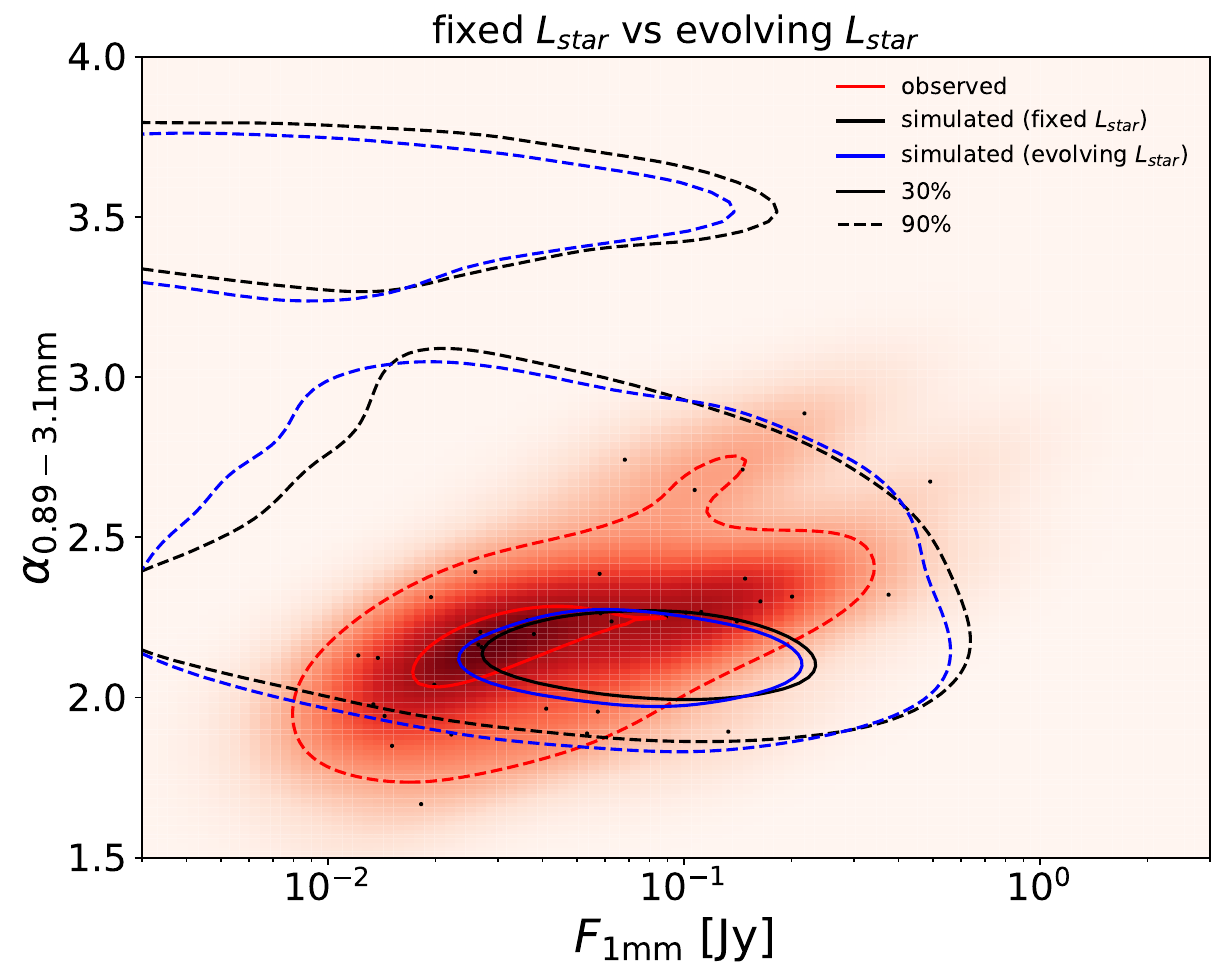}
    \caption{Spectral index distribution sub-structured disks, for the entire parameter space of initial conditions (Table \ref{table:2}) for the scenario in which we keep $L_{star}$ fixed during the disk evolution (black lines) and the scenario in which $L_{star}$ evolve by time. Heatmap of the observed disks with the black dots representing each single observed disk. The black, blue and red lines refer to the simulated results obtained for fixed $L_{star}$ (black), simulated results obtained for evolving $L_{star}$ (blue) and the observational results (red). In particular, the continuous lines encompass the $30\%$ of the cumulative sum of the disks produced from the simulations or observed. The dashed lines encompass the $90\%$ instead.}
    \label{Lstar_comparison}
\end{figure*}
In this appendix we show the comparison between the spectral index simulated distribution obtained for a scenario in which we keep the luminosity of the hosting star fixed and the scenario in which we let $L_{\star}$ evolve by time.
\FloatBarrier
\section{Initial parameter distributions}
\label{appendix:initial_conditions}
In this appendix we show the spectral index, size-luminosity and initial parameters distributions for the entire parameter space of the initial conditions (see Table \ref{table:2}). 
\begin{figure*}[h]
    \centering
    \includegraphics[scale=0.6]{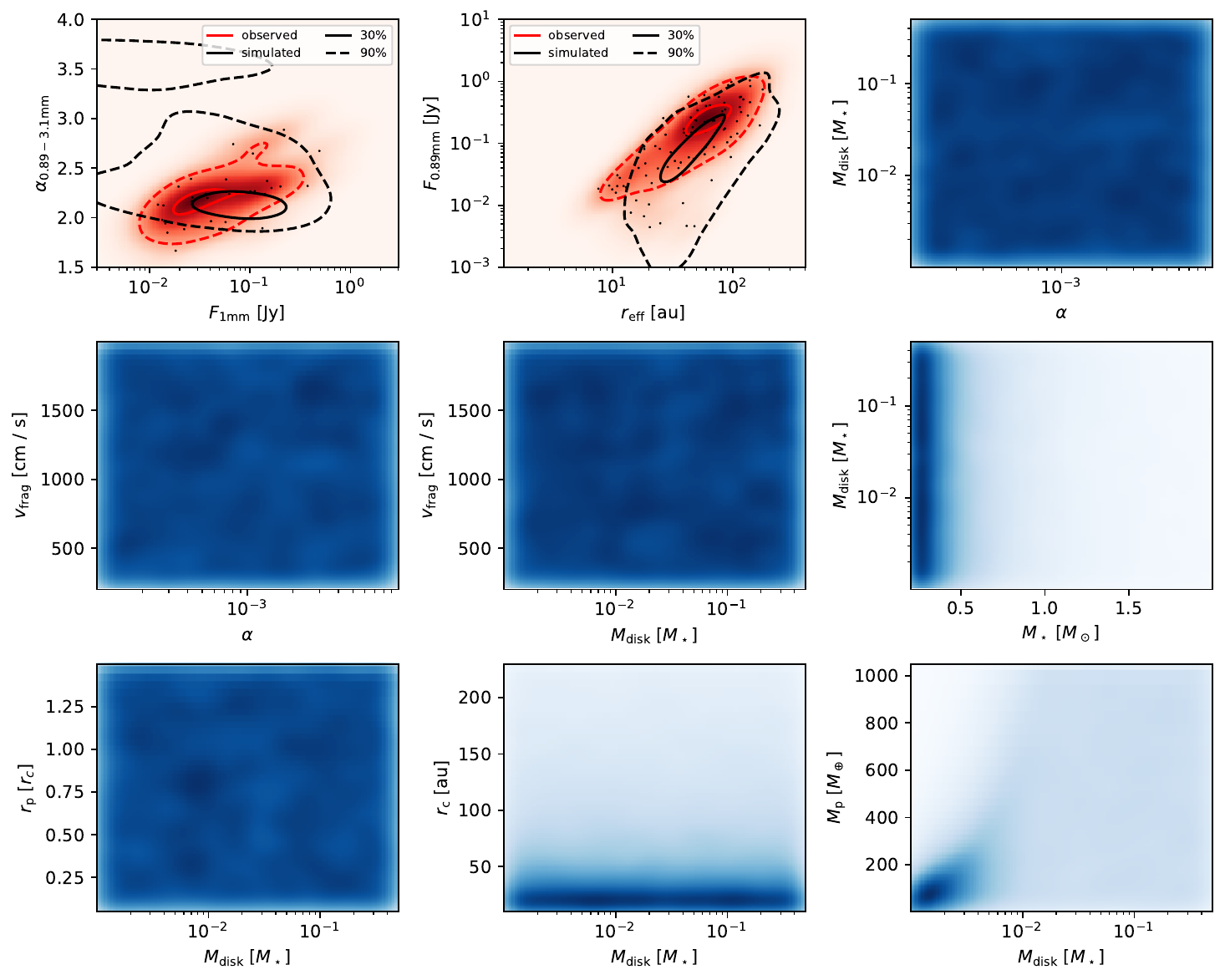}
    \caption{Spectral index, size-luminosity and initial parameters distributions for the entire parameter space of the initial conditions. We compare our simulated spectral index distributions to the observed sample adopted in \cite{tazzari2021first}. The latter is a collection of disks from Lupus region, detected at 0.89mm \citep{ansdell2016alma} and 3.1mm \citep{tazzari2021first}, and Taurus and Ophiucus star-forming regions \citep{ricci2010dust_b,ricci2010dust_a}. We compare our simulated SLR to the observed sample reported in \cite{Andrews2018a}.}
    \label{initialcond_noobscuts}
\end{figure*}
\FloatBarrier
\section{Size-luminosity diagram analysis}
\label{appendix:slr_analysis}
In this appendix we show the spectral index, size-luminosity and initial parameters distributions selecting disks populating two different regions in the size-luminosity diagram. 
\begin{figure}[htb]
    \centering
    \includegraphics[scale=0.35]{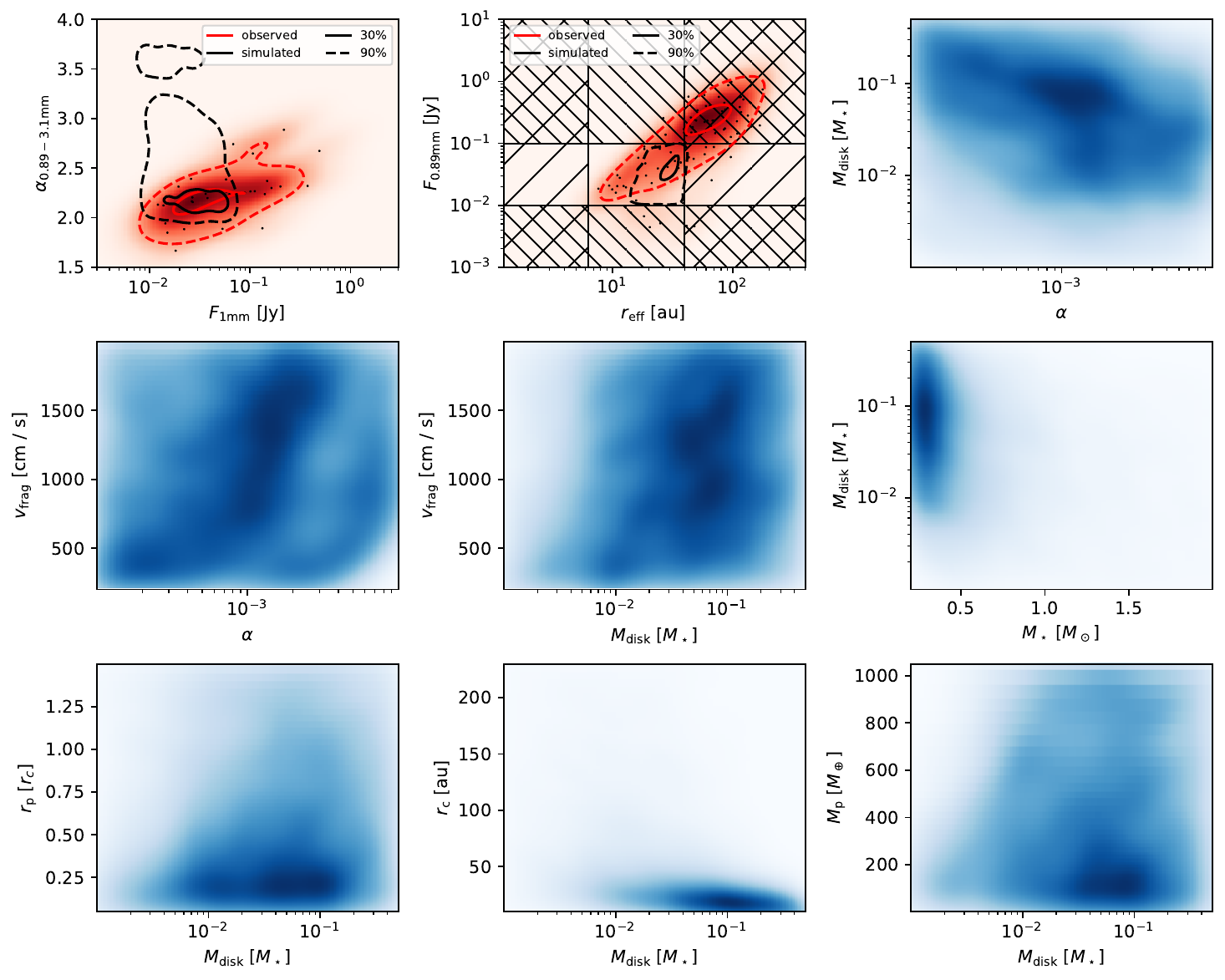}
    \caption{Spectral index, size-luminosity and initial parameters distributions selecting disks with $0.8\leqslant log\ r_{eff}[au]\leqslant1.6$ and $-2\leqslant log\ F_{0.89mm}[Jy]\leqslant-1$.}
    \label{slr_bottom}
\end{figure}

\begin{figure}[htb]
    \centering
    \includegraphics[scale=0.35]{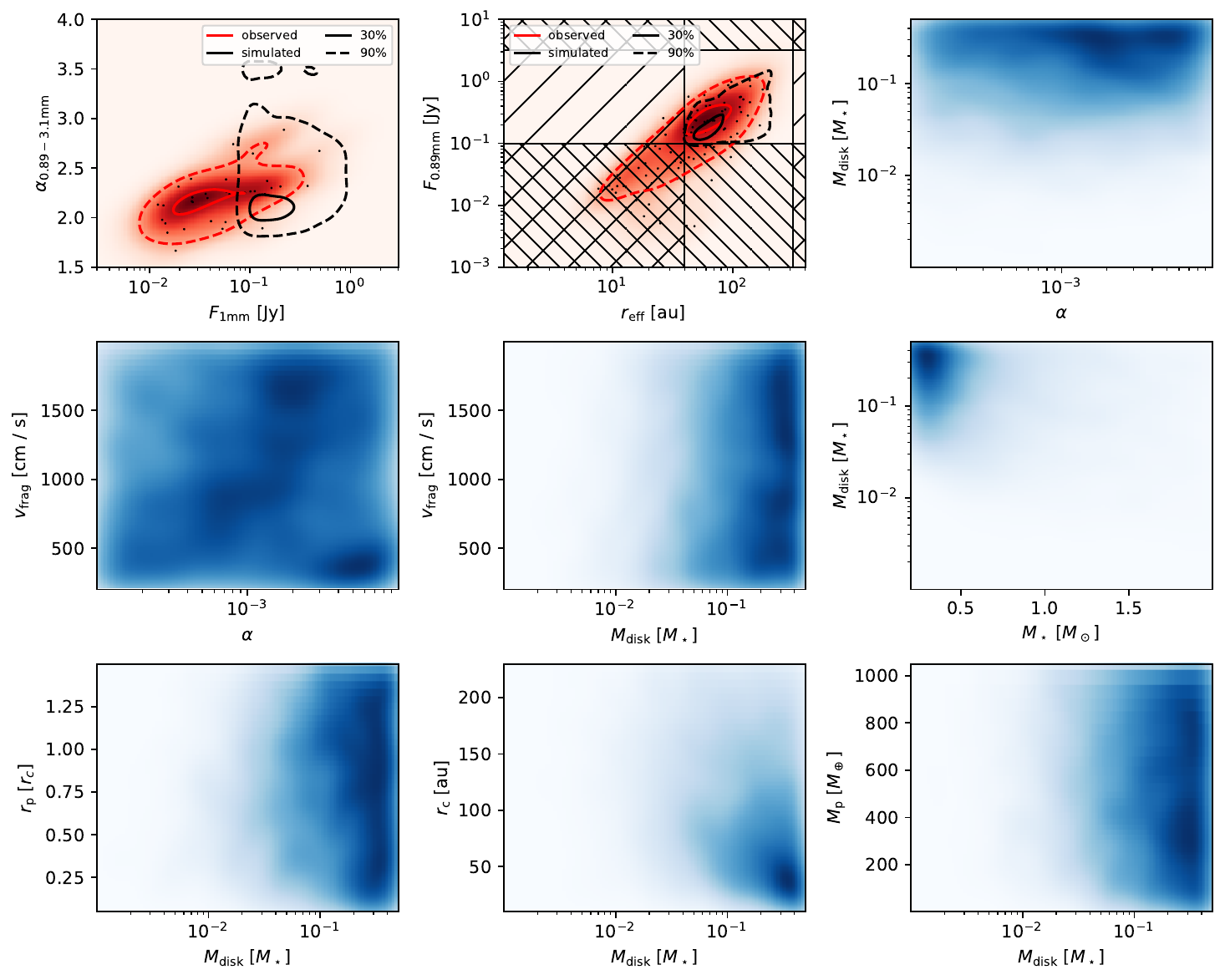}
    \caption{Spectral index, size-luminosity and initial parameters distributions selecting disks with $1.6\leqslant log\ r_{eff}[au]\leqslant2.5$ and $-1\leqslant log\ F_{0.89mm}[Jy]\leqslant0.5$.}
    \label{slr_top}
\end{figure}
\FloatBarrier
\section{DSHARP opacities for different \% of grain porosity}\label{appendix:dsharp_porosities}
In this appendix we show the comparison between the spectral index simulated distribution obtained for DHSARP opacities \citep{Birnstiel2018} for different \% of grain porosity.
\begin{figure*}[!htb]
    \centering
    \includegraphics[scale=0.45]{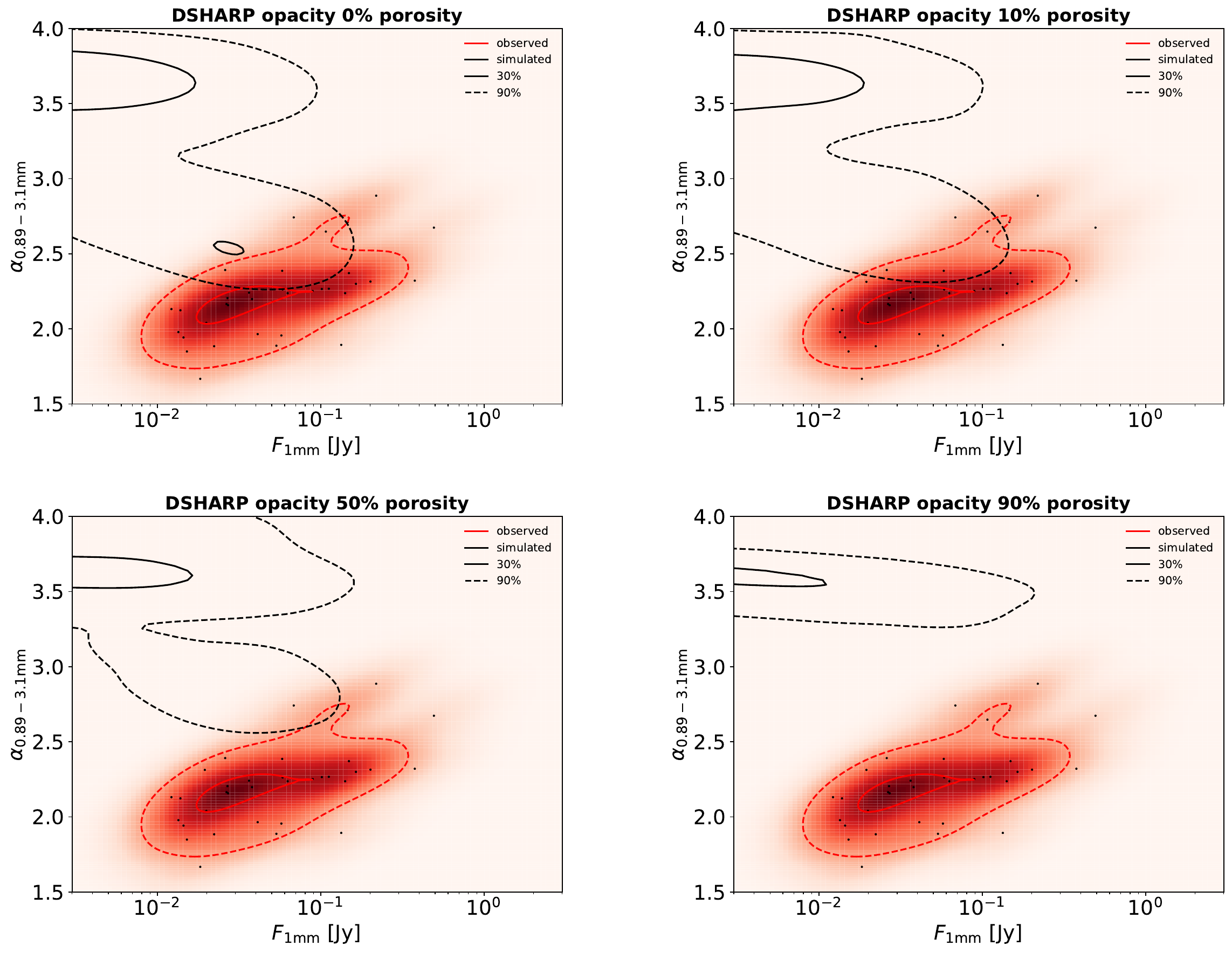}
    \caption{Spectral index distribution sub-structured disks, for the entire parameter space of initial conditions (Table \ref{table:2}) for DSHARP opacities \citep{Birnstiel2018} for four different \% of grain porosity. Heatmap of the observed disks with the black dots representing each observed disk. The black and red lines refer to the simulated results and the observational results respectively. In particular, the continuous lines encompass the $30\%$ of the cumulative sum of the disks produced from the simulations or observed. The dashed lines encompass the $90\%$ instead.}
    \label{spectral_dsharp_porosity}
\end{figure*}
\end{appendix}

\end{document}